\documentclass[]{aa}
\usepackage{epsfig}
\begin{document}
\def\gsim{\vcenter{\hbox{$>$}\offinterlineskip\hbox{$\sim$}}}
\def\lsim{\vcenter{\hbox{$<$}\offinterlineskip\hbox{$\sim$}}}
\title{An empirical formula for the mass-loss rates of dust-enshrouded red
       supergiants and oxygen-rich Asymptotic Giant Branch stars\thanks{based
       on observations obtained at the European Southern Observatory (La
       Silla, Chile)}}
\author{Jacco Th. van Loon\inst{1}, Maria-Rosa L. Cioni\inst{2,3}, Albert A.
        Zijlstra\inst{4}, Cecile Loup\inst{5}}
\institute{Astrophysics Group, School of Physical \& Geographical Sciences,
           Keele University, Staffordshire ST5 5BG, UK
      \and European Southern Observatory, Karl-Schwarzschild Stra{\ss}e 2,
           D-85748 Garching bei M\"{u}nchen, Germany
      \and Institute for Astronomy, University of Edinburgh, Royal
           Observatory, Blackford Hill, Edinburgh EH9 3HJ, UK
      \and School of Physics and Astronomy, University of Manchester,
           Sackville Street, P.O.Box 88, Manchester M60 1QD, UK
      \and Institut d'Astrophysique de Paris, 98bis Boulevard Arago, F-75014
           Paris, France
}
\offprints{\email{jacco@astro.keele.ac.uk}}
\date{Received date; Accepted date}
\titlerunning{Mass-loss rates of red supergiants and AGB stars}
\authorrunning{van Loon, Cioni, Zijlstra, Loup}
\abstract{
We present an empirical determination of the mass-loss rate as a function of
stellar luminosity and effective temperature, for oxygen-rich dust-enshrouded
Asymptotic Giant Branch stars and red supergiants. To this aim we obtained
optical spectra of a sample of dust-enshrouded red giants in the Large
Magellanic Cloud, which we complemented with spectroscopic and infrared
photometric data from the literature. Two of these turned out to be hot
emission-line stars, of which one is a definite B[e] star. The mass-loss rates
were measured through modelling of the spectral energy distributions. We thus
obtain the mass-loss rate formula $\log \dot{M} = -5.65 + 1.05 \log ( L /
10,000\, {\rm L}_\odot ) -6.3 \log ( T_{\rm eff} / 3500\, {\rm K} )$, valid
for dust-enshrouded red supergiants and oxygen-rich AGB stars. Despite the low
metallicity of the LMC, both AGB stars and red supergiants are found at late
spectral types. A comparison with galactic AGB stars and red supergiants shows
excellent agreement between the mass-loss rate as predicted by our formula and
that derived from the 60\,$\mu$m flux density for dust-enshrouded objects, but
not for optically bright objects. We discuss the possible implications of this
for the mass-loss mechanism.
}
\maketitle

\section{Introduction}

Stellar death results in the injection of chemically enriched mass, energy and
momentum into the interstellar medium. It also results in baryonic mass being
locked up forever into degenerate objects (white dwarves, neutron stars and
black holes). Together, these processes drive galactic chemical evolution and
star formation.

Intermediate-mass stars, with Main-Sequence masses of $M_{\rm ZAMS}\sim$1--8
M$_\odot$, become Asymptotic Giant Branch (AGB) stars before ending as a white
dwarf on timescales from only $t\simeq$30 Myr up to as much as 10 Gyr.
Luminous, $L_{\rm AGB}\sim10^4$ L$_\odot$, and cool, $T_{\rm eff,
AGB}\sim2500$--3500 K, stars on the upper slopes of the AGB experience intense
mass loss, losing up to 80 per cent of their original mass at rates of
$\dot{M}_{\rm AGB}\sim10^{-6}$--$10^{-4}$ M$_\odot$ yr$^{-1}$ (Wood, Bessell
\& Fox 1983; Wood et al.\ 1992). Their luminosity is normally generated in a
hydrogen-burning shell, isolated from the degenerate carbon-oxygen core and
the overlying convective envelope, but episodically the temperature becomes
sufficiently high to temporarily ignite a helium-burning shell (thermal
pulses). This leads to penetration of the convective envelope, allowing for
nuclear processed material to be transported to the stellar surface (e.g.,
carbon and s-process elements such as lithium and technetium). AGB stars
within a certain mass range can thus be turned into carbon stars when the
photospheric carbon-to-oxygen abundance ratio exceeds unity (Bessell, Wood \&
Lloyd Evans 1983).

Massive stars also experience intense mass loss before they end through a
supernova, leaving a neutron star or black hole (or deflagrating entirely) on
timescales of $t\sim$1 to 30 Myr. Of these the vast majority, stars with
$M_{\rm ZAMS}\sim$8--30 M$_\odot$ pass through a red supergiant (RSG) stage,
when they both are very luminous, $L_{\rm RSG}\sim10^5$ L$_\odot$, and become
cool, $T_{\rm eff, RSG}\sim3500$--4000 K (Massey \& Olsen 2003). As red
supergiants they loose mass at rates of $\dot{M}\sim10^{-5}$--$10^{-3}$
M$_\odot$ yr$^{-1}$, but there is a lot of uncertainty about the total amount
of mass lost during the RSG phase, and therefore about the exact nature of the
immediate supernova progenitors (Filippenko 1997).

AGB and RSG mass loss is essentially driven by the same mechanism: radiation
pressure on dust grains which condense in the elevated atmospheres of radially
pulsating giant stars and drag the gas along with it through mutual
collisions. It seems therefore logical that the mass-loss rate would depend on
the stellar luminosity (radiation pressure) and temperature (to allow dust to
form). Indeed, both theory and observations confirm that this is the case.
However, even the most sophisticated, hydrodynamical models suffer from a
range of assumptions and are so far only presented for carbon stars, at an
initial solar metallicity (e.g., Wachter et al.\ 2002; Sandin \& H\"{o}fner
2003). Mass-loss rates have been measured for a sample of AGB stars and red
supergiants in the Large Magellanic Cloud (LMC) by van Loon et al.\ (1999),
who demonstrate that the mass-loss rate increases with increasing luminosity,
and for increasingly late spectral type for oxygen-rich (M-type) AGB stars and
red supergiants, but they do not quantify these dependencies.

We here present a recipe for the mass-loss rate as a function of luminosity
and stellar effective temperature, $\dot{M}(L,T_{\rm eff})$, for oxygen-rich
AGB stars and red supergiants in the LMC, which undergo strong regular
pulsations (Whitelock et al.\ 2003) and are enshrouded in dusty envelopes.
This is the first empirical relationship of its kind, that can be readily
applied in models for stellar and galactic evolution. Although much work has
been done on the mass-loss rates of cool giants in globular clusters and in
the solar vicinity (e.g., Reimers 1975a,b; Jura \& Kleinmann 1992; Catelan
2000), these studies concentrate on first ascent RGB stars and low-mass AGB
stars with moderate mass-loss rates. Our formula is the first attempt to
quantify the relationship between the measured values for the mass-loss rate,
bolometric luminosity and stellar effective temperature for dusty cool giants.
Valid well into the superwind regime of extremely high mass-loss rates, our
formula provides a more direct alternative to the commonly used
extra-polations of relationships which were derived for less massive and less
evolved red giants.

As part of this investigation, we obtained optical spectra for a number of
highly obscured objects for which no such information was available, to
estimate their stellar effective temperatures. We compiled infrared (IR)
photometry to sample the spectral energy distributions which were then
modelled with a dust radiative transfer code in order to derive the mass-loss
rates. We also present a more qualitative analysis of a small sample of carbon
stars and (M)S-type stars with a carbon-to-oxygen ratio slightly less than
unity, as well as a comparison with galactic AGB stars and red supergiants.

\section{Observational data}

\subsection{Spectroscopy}

%
%
\begin{table}
\caption[]{Spectroscopic targets, listed in order of increasing Right
Ascension (J2000 coordinates). The last column refers to the observing run,
where ``D5'' = DFOSC 1995, ``E8'' = EMMI 1998 and ``E0'' = EMMI 2000.}
\begin{tabular}{lccr}
\hline\hline
Object                &
RA                    &
Dec                   &
\llap{R}un            \\
\hline
IRAS\,04407$-$7000    &
4 40 28.5             &
$-$69 55 14           &
E8                    \\
IRAS\,04498$-$6842    &
4 49 41.5             &
$-$68 37 52           &
E8                    \\
IRAS\,04530$-$6916    &
4 52 45.7             &
$-$69 11 50           &
E8                    \\
WOH\,G64              &
4 55 10.5             &
$-$68 20 30           &
D5                    \\
IRAS\,05003$-$6712    &
5 00 19.0             &
$-$67 07 58           &
E8                    \\
SP77\,28-10           &
5 00 23.0             &
$-$65 59 59           &
E0                    \\
DCMC\,J050024.95$-$710130.\rlap{7} &
5 00 25.0             &
$-$71 01 31           &
E0                    \\
SHV\,0500533$-$690617 &
5 00 38.2             &
$-$69 02 01           &
E0                    \\
SHV\,0504018$-$690157 &
5 03 46.8             &
$-$68 57 54           &
E0                    \\
SHV\,0504353$-$712622 &
5 03 56.0             &
$-$71 22 22           &
E0                    \\
SHV\,0504266$-$691145 &
5 04 10.1             &
$-$69 07 43           &
E0                    \\
DCMC\,J050615.14$-$720657.\rlap{1} &
5 06 15.1             &
$-$72 06 57           &
E0                    \\
DCMC\,J050659.94$-$634746.\rlap{7} &
5 06 59.9             &
$-$63 47 47           &
E0                    \\
GRV\,0507$-$6616      &
5 07 10.3             &
$-$66 12 43           &
E0                    \\
DCMC\,J050738.74$-$733233.\rlap{4} &
5 07 38.7             &
$-$73 32 33           &
E0                    \\
IRAS\,05128$-$6455    &
5 13 04.6             &
$-$64 51 40           &
E8                    \\
HV\,916               &
5 14 49.7             &
$-$67 27 20           &
D5                    \\
GRV\,0519$-$6700      &
5 19 42.0             &
$-$66 57 49           &
D5                    \\
HV\,2561              &
5 28 28.9             &
$-$68 07 08           &
D5                    \\
IRAS\,05294$-$7104    &
5 28 48.2             &
$-$71 02 29           &
E8                    \\
IRAS\,05289$-$6617    &
5 29 02.4             &
$-$66 15 28           &
E8                    \\
SP77\,46-44           &
5 29 42.2             &
$-$68 57 17           &
D5                    \\
HV\,986               &
5 31 09.3             &
$-$67 25 55           &
D5                    \\
LI-LMC\,1522          &
5 40 13.3             &
$-$69 22 47           &
E8                    \\
\hline
\end{tabular}
\end{table}

Spectra of (candidate) dust-enshrouded AGB stars and red supergiants were
obtained on three occasions (1995, 1998 and 2000). DFOSC at the 1.5m Danish
telescope at La Silla, Chile, was used from 8 to 11 December 1995 to obtain
low-resolution ($R\sim900$) long-slit spectra between $\sim$0.41 and 0.68
$\mu$m with grism \#7 and a $1.5^{\prime\prime}$ slit width. EMMI at the ESO
3.5m NTT at La Silla, Chile, was used on 24 December 1998 and on 15 \& 16
January 2000, to obtain low-resolution ($R\sim500$) long-slit spectra between
$\sim$0.61 and 1.04 $\mu$m with grism \#4 and a $1.5^{\prime\prime}$ slit
width. Integration times ranged between 5 and 30 minutes.

The data were reduced using standard procedures in ESO-Midas. The frames were
corrected for the electronic offset (bias) and the relative pixel response
(flatfield). Wavelength calibration was performed relative to He+Ar or He+Ne
lamp spectra. The sky-subtracted spectra were corrected for the wavelength
dependence of the instrumental response as measured from the spectrum of a
standard star, and for atmospheric continuum extinction.

The targets (Table 1) were chosen from samples of mid-IR sources with known
optically visible counterparts (dominating the DFOSC 1995 run) --- mainly from
Loup et al.\ (1997), bright mid-IR sources associated with heavily
dust-enshrouded objects (mostly during the EMMI 1998 run) --- cf.\ Zijlstra et
al.\ (1996), and red objects that were likely to be carbon stars (EMMI 2000
run). The latter were selected from Cioni et al.\ (2000) to add to the
previous samples that turned out to be predominantly oxygen-rich. Several
galactic M giants were also observed for reference (see Appendix A).

\subsection{Infrared photometry}

%
%
\begin{table*}
\caption[]{Spectral classification and near-IR magnitudes (wavelengths in
$\mu$m).}
\begin{tabular}{lll|rrr|rrr}
\hline\hline
Object                      &
\multicolumn{2}{c|}{Spectral type} &
\multicolumn{3}{c|}{DENIS}  &
\multicolumn{3}{c}{2MASS}   \\
                            &
Here                        &
Previous                    &
0.79                        &
1.23                        &
2.15                        &
1.24                        &
1.66                        &
2.16                        \\
\hline
{\it Stars of type M} & & & & & & & & \\
HV\,12501                   &
                            &
M1.5                        &
10.08                       &
8.62                        &
7.51                        &
8.80                        &
8.03                        &
7.70                        \\
HV\,2360                    &
                            &
M2\,Ia                      &
10.53                       &
                            &
7.45                        &
8.78                        &
7.98                        &
7.59                        \\
HV\,2446                    &
                            &
M5e                         &
14.38                       &
10.56                       &
8.97                        &
10.67                       &
9.81                        &
9.36                        \\
HV\,2561                    &
M1.5                        &
M0\,Ia                      &
10.09                       &
8.26                        &
6.79                        &
8.61                        &
7.80                        &
7.48                        \\
HV\,5870                    &
                            &
M4.5/5                      &
10.95                       &
                            &
7.70                        &
9.10                        &
8.26                        &
7.90                        \\
HV\,888                     &
                            &
M4\,Ia                      &
9.66                        &
7.88                        &
                            &
8.01                        &
7.19                        &
6.78                        \\
HV\,916                     &
M2.5                        &
M3\,Iab                     &
10.47                       &
                            &
7.34                        &
8.64                        &
7.78                        &
7.42                        \\
HV\,986                     &
M2                          &
                            &
                            &
                            &
                            &
8.80                        &
7.95                        &
7.63                        \\
HV\,996                     &
                            &
M4\,Iab                     &
10.71                       &
8.82                        &
                            &
8.99                        &
8.16                        &
7.64                        \\
IRAS\,04407$-$7000          &
M7.5                        &
                            &
13.73                       &
10.09                       &
8.11                        &
10.66                       &
9.49                        &
8.69                        \\
IRAS\,04498$-$6842          &
M10                         &
                            &
14.96                       &
10.67                       &
8.40                        &
9.13                        &
8.05                        &
7.49                        \\
IRAS\,04509$-$6922          &
                            &
M10                         &
                            &
                            &
                            &
9.87                        &
8.67                        &
7.93                        \\
IRAS\,04516$-$6902          &
                            &
M9                          &
17.39                       &
11.35                       &
                            &
9.93                        &
8.62                        &
7.91                        \\
IRAS\,05003$-$6712          &
M9                          &
                            &
18.29                       &
13.17                       &
10.32                       &
12.04                       &
10.46                       &
9.32                        \\
IRAS\,05128$-$6455          &
M9                          &
                            &
17.87                       &
13.43                       &
                            &
14.55                       &
12.83                       &
11.28                       \\
IRAS\,05294$-$7104          &
M8                          &
                            &
                            &
12.87                       &
9.67                        &
12.99                       &
11.19                       &
9.88                        \\
SHV\,0522023$-$701242       &
                            &
M3                          &
16.67                       &
13.29                       &
                            &
12.63                       &
11.74                       &
11.31                       \\
SHV\,0530323$-$702216       &
                            &
M6                          &
14.66                       &
11.49                       &
9.86                        &
11.79                       &
10.89                       &
10.39                       \\
SP77\,30-6                  &
                            &
M8                          &
14.61                       &
11.06                       &
9.36                        &
11.14                       &
10.28                       &
9.55                        \\
SP77\,46-44                 &
M1.5                        &
M1\,Ia                      &
9.53                        &
7.83                        &
6.38                        &
7.92                        &
7.19                        &
6.89                        \\
WOH\,G64                    &
M7.5e                       &
M7.5                        &
13.04                       &
9.44                        &
                            &
9.25                        &
7.74                        &
6.85                        \\
WOH\,SG374                  &
                            &
M6                          &
                            &
                            &
                            &
9.80                        &
9.18                        &
8.67                        \\
\hline
{\it Stars of type MS and S} & & & & & & & & \\
GRV\,0519$-$6700            &
S5,2                        &
                            &
14.81                       &
13.04                       &
10.83                       &
12.17                       &
11.17                       &
10.59                       \\
HV\,12070                   &
                            &
MS3/9                       &
14.37                       &
10.43                       &
9.02                        &
10.61                       &
9.75                        &
9.20                        \\
SHV\,0524565$-$694559       &
                            &
MS5                         &
14.26                       &
12.06                       &
10.24                       &
11.60                       &
10.67                       &
10.27                       \\
\hline
{\it Stars of type C} & & & & & & & & \\
DCMC\,J050024.95$-$710130.7 &
C9,5+                       &
                            &
15.31                       &
13.02                       &
10.33                       &
12.46                       &
11.17                       &
10.10                       \\
DCMC\,J050615.14$-$720657.1 &
C6,3                        &
                            &
15.18                       &
13.06                       &
10.25                       &
13.13                       &
11.64                       &
10.35                       \\
DCMC\,J050659.94$-$634746.7 &
C9,5                        &
                            &
14.35                       &
12.35                       &
10.12                       &
12.45                       &
11.08                       &
10.14                       \\
DCMC\,J050738.74$-$733233.4 &
C1,?                        &
                            &
14.45                       &
12.17                       &
9.73                        &
13.12                       &
11.60                       &
10.43                       \\
GRV\,0507$-$6616            &
C8,4                        &
                            &
15.45                       &
13.35                       &
10.62                       &
12.99                       &
11.53                       &
10.50                       \\
IRAS\,05289$-$6617          &
C10,1                       &
                            &
15.99                       &
14.32                       &
12.77                       &
14.77                       &
13.62                       &
13.02                       \\
SHV\,0500533$-$690617       &
C8,5+                       &
                            &
16.12                       &
13.54                       &
11.13                       &
13.92                       &
12.39                       &
11.44                       \\
SHV\,0504018$-$690157       &
C5,4                        &
                            &
15.78                       &
13.53                       &
10.54                       &
13.59                       &
11.94                       &
10.69                       \\
SHV\,0504266$-$691145       &
C8,4                        &
                            &
15.23                       &
13.05                       &
10.67                       &
14.38                       &
12.72                       &
11.47                       \\
SHV\,0504353$-$712622       &
C7,5                        &
                            &
14.87                       &
12.78                       &
10.48                       &
12.37                       &
11.13                       &
10.41                       \\
SP77\,28-10                 &
C9,3                        &
                            &
15.85                       &
13.48                       &
10.71                       &
12.99                       &
11.56                       &
10.51                       \\
\hline
{\it Emission-line stars} & & & & & & & & \\
IRAS\,04530$-$6916          &
em.                         &
                            &
15.72                       &
13.47                       &
9.61                        &
13.94                       &
11.86                       &
9.96                        \\
LI-LMC\,1522                &
em.                         &
                            &
10.93                       &
10.31                       &
8.43                        &
10.46                       &
9.85                        &
8.60                        \\
\hline
\end{tabular}
\end{table*}

%
%
\begin{table*}
\caption[]{Mid-IR flux densities (in Jy, with the wavelengths in $\mu$m). The
column ``s'' indicates the availability of an ISO spectrum (``y''). Values
accompanied by a colon are uncertain, whilst question marks indicate
detections that are suspect.}
\begin{tabular}{l|rrrr|rrr|rrrr}
\hline\hline
Object                      &
\multicolumn{4}{c|}{MSX}    &
\multicolumn{3}{c|}{ISO}    &
\multicolumn{4}{c}{IRAS}    \\
                            &
8.28                        &
12.13                       &
14.65                       &
21.34                       &
12                          &
25                          &
s                           &
12                          &
25                          &
60                          &
100                         \\
\hline
{\it Stars of type M} & & & & & & & & & & \\
HV\,12501                   &
0.108                       &
$<$0.05                     &
$<$0.08                     &
$<$0.4                      &
0.19                        &
0.07                        &
                            &
0.2\rlap{0}                 &
0.0\rlap{8}                 &
$<$0.2                      &
                            \\
HV\,2360                    &
0.321                       &
0.41                        &
0.25                        &
$<$0.8                      &
0.33                        &
0.14                        &
                            &
0.3\rlap{8}                 &
0.3\rlap{5}                 &
0.4\rlap{:}                 &
                            \\
HV\,2446                    &
0.033\rlap{:}               &
                            &
                            &
                            &
0.07                        &
0.04                        &
y                           &
0.0\rlap{5}                 &
0.0\rlap{2}                 &
$<$0.2                      &
                            \\
HV\,2561                    &
0.245                       &
0.18\rlap{:}                &
0.15\rlap{:}                &
0.5\rlap{:}                 &
                            &
                            &
                            &
0.2\rlap{8}                 &
0.2\rlap{2}                 &
$<$1.0                      &
                            \\
HV\,5870                    &
0.144                       &
$<$0.12                     &
$<$0.06                     &
$<$0.5                      &
0.27                        &
0.09                        &
                            &
0.3\rlap{0}                 &
0.1\rlap{7}                 &
$<$5.0                      &
                            \\
HV\,888                     &
0.430                       &
0.51                        &
0.17                        &
$<$0.2                      &
0.71                        &
0.20                        &
y                           &
0.5\rlap{8}                 &
0.2\rlap{9}                 &
$<$4.0                      &
                            \\
HV\,916                     &
0.267                       &
0.23\rlap{:}                &
0.27\rlap{:}                &
$<$0.4                      &
0.38                        &
0.18                        &
                            &
0.4\rlap{4}                 &
0.2\rlap{3}                 &
$<$2.0                      &
                            \\
HV\,986                     &
0.076                       &
$<$0.19                     &
$<$0.13                     &
$<$0.5                      &
                            &
                            &
                            &
0.1\rlap{1}                 &
0.0\rlap{6:}                &
$<$0.6                      &
                            \\
HV\,996                     &
0.457                       &
0.67                        &
0.27                        &
0.3\rlap{:}                 &
0.60                        &
0.36                        &
y                           &
0.7\rlap{1}                 &
0.5\rlap{3}                 &
$<$0.5                      &
                            \\
IRAS\,04407$-$7000          &
0.424                       &
0.46                        &
0.51                        &
0.7                         &
0.96                        &
0.58                        &
                            &
0.7\rlap{6}                 &
0.7\rlap{6}                 &
0.1                         &
                            \\
IRAS\,04498$-$6842          &
0.321                       &
0.64                        &
0.73                        &
$<$0.5                      &
0.49                        &
0.22                        &
                            &
1.3\rlap{3}                 &
0.8\rlap{9}                 &
$<$0.2                      &
                            \\
IRAS\,04509$-$6922          &
0.181                       &
$<$0.20                     &
0.34\rlap{:}                &
$<$0.3                      &
0.30                        &
0.20                        &
                            &
0.8\rlap{9}                 &
0.8\rlap{6}                 &
$<$2.0                      &
                            \\
IRAS\,04516$-$6902          &
0.318                       &
0.21                        &
$<$0.18                     &
$<$0.4                      &
0.85                        &
0.38                        &
                            &
0.8\rlap{6}                 &
0.5\rlap{5}                 &
0.4\rlap{:}                 &
                            \\
IRAS\,05003$-$6712          &
0.155                       &
0.34\rlap{:}                &
0.17\rlap{:}                &
$<$0.1                      &
0.36                        &
0.21                        &
y                           &
0.4\rlap{3}                 &
0.3\rlap{3}                 &
0.1\rlap{:}                 &
                            \\
IRAS\,05128$-$6455          &
0.196                       &
$<$0.28                     &
0.22\rlap{:}                &
$<$0.2                      &
0.23                        &
0.06                        &
y                           &
0.2\rlap{3}                 &
0.2\rlap{4}                 &
0.1                         &
                            \\
IRAS\,05294$-$7104          &
0.177                       &
0.41\rlap{:}                &
$<$0.22                     &
$<$0.4                      &
0.68                        &
0.39                        &
                            &
0.6\rlap{9}                 &
0.5\rlap{6}                 &
$<$3.0                      &
                            \\
SHV\,0522023$-$701242       &
$<$0.010                    &
                            &
                            &
                            &
0.00\rlap{1:}               &
$<$0.02                     &
                            &
$<$0.1\rlap{0}              &
$<$0.0\rlap{4}              &
0.4\rlap{:}                 &
                            \\
SHV\,0530323$-$702216       &
0.011\rlap{:}               &
                            &
                            &
                            &
0.00\rlap{8}                &
0.00\rlap{8:}               &
                            &
$<$0.0\rlap{4}              &
$<$0.0\rlap{4}              &
0.4                         &
                            \\
SP77\,30-6                  &
0.089                       &
$<$0.12                     &
$<$0.15                     &
$<$0.5                      &
0.14                        &
0.08                        &
y                           &
0.2\rlap{6}                 &
0.1\rlap{3}                 &
0.1\rlap{:}                 &
                            \\
SP77\,46-44                 &
0.192                       &
0.17\rlap{:}                &
$<$0.21                     &
$<$0.8                      &
                            &
                            &
                            &
0.2\rlap{6}                 &
0.1\rlap{8}                 &
$<$0.2                      &
                            \\
WOH\,G64                    &
6.220                       &
8.73                        &
9.71                        &
11.6                        &
\llap{1}2.11                &
\llap{1}3.79                &
y                           &
8.4\rlap{5}                 &
13.5\rlap{3}                &
2.2                         &
\llap{$<$}16.0              \\
WOH\,SG374                  &
0.254                       &
0.44                        &
0.39\rlap{:}                &
$<$0.2                      &
0.49                        &
0.19                        &
                            &
0.3\rlap{7}                 &
0.3\rlap{8}                 &
0.2\rlap{:}                 &
                            \\
\hline
{\it Stars of type MS and S} & & & & & & & & & & \\
GRV\,0519$-$6700            &
$<$0.023                    &
                            &
                            &
                            &
0.00\rlap{4}                &
                            &
                            &
$<$0.0\rlap{6}              &
$<$0.0\rlap{3}              &
0.0\rlap{8:}                &
                            \\
HV\,12070                   &
0.054                       &
                            &
                            &
                            &
0.04\rlap{3}                &
$<$0.02                     &
y                           &
0.0\rlap{6}                 &
0.0\rlap{3}                 &
0.1\rlap{:}                 &
                            \\
SHV\,0524565$-$694559       &
$<$0.008                    &
                            &
                            &
                            &
0.00\rlap{3}                &
                            &
                            &
$<$0.1\rlap{4}              &
$<$0.0\rlap{7}              &
$<$1.0                      &
                            \\
\hline
{\it Stars of type C} & & & & & & & & & & \\
DCMC\,J050024.95$-$710130.7 &
0.074                       &
                            &
                            &
                            &
                            &
                            &
                            &
0.0\rlap{7?}                &
0.0\rlap{5?}                &
$<$0.1                      &
                            \\
DCMC\,J050615.14$-$720657.1 &
0.022\rlap{:}               &
                            &
                            &
                            &
                            &
                            &
                            &
0.1\rlap{0}                 &
0.0\rlap{4}                 &
$<$0.2                      &
                            \\
DCMC\,J050659.94$-$634746.7 &
$<$0.020                    &
                            &
                            &
                            &
                            &
                            &
                            &
$<$0.1\rlap{0}              &
$<$0.1\rlap{0}              &
$<$0.2                      &
                            \\
DCMC\,J050738.74$-$733233.4 &
0.041                       &
                            &
                            &
                            &
                            &
                            &
                            &
0.0\rlap{7}                 &
0.0\rlap{2:}                &
$<$0.2                      &
                            \\
GRV\,0507$-$6616            &
0.022\rlap{:}               &
                            &
                            &
                            &
                            &
                            &
                            &
$<$0.0\rlap{4}              &
0.0\rlap{5:}                &
$<$0.2                      &
                            \\
IRAS\,05289$-$6617          &
0.064                       &
$<$0.41                     &
0.10\rlap{:}                &
0.2\rlap{:}                 &
0.16                        &
0.20                        &
y                           &
0.1\rlap{6}                 &
0.3\rlap{9}                 &
0.3                         &
                            \\
SHV\,0500533$-$690617       &
0.034\rlap{:}               &
                            &
                            &
                            &
                            &
                            &
                            &
0.0\rlap{5?}                &
$<$0.0\rlap{8}              &
$<$0.6                      &
                            \\
SHV\,0504018$-$690157       &
0.045\rlap{:}               &
                            &
                            &
                            &
                            &
                            &
                            &
0.0\rlap{3}                 &
$<$0.1\rlap{3}              &
$<$0.6                      &
                            \\
SHV\,0504266$-$691145       &
0.029\rlap{:}               &
                            &
                            &
                            &
                            &
                            &
                            &
0.0\rlap{2:}                &
$<$0.1\rlap{0}              &
$<$2.0                      &
                            \\
SHV\,0504353$-$712622       &
$<$0.009                    &
                            &
                            &
                            &
                            &
                            &
                            &
0.0\rlap{2?}                &
0.0\rlap{3?}                &
0.5\rlap{?}                 &
                            \\
SP77\,28-10                 &
0.031                       &
                            &
                            &
                            &
                            &
                            &
                            &
$<$0.1\rlap{5}              &
$<$0.1\rlap{1}              &
$<$0.9                      &
                            \\
\hline
{\it Emission-line stars} & & & & & & & & & & \\
IRAS\,04530$-$6916          &
1.144                       &
1.93                        &
1.72                        &
2.5                         &
                            &
                            &
                            &
2.0\rlap{7}                 &
5.0\rlap{9}                 &
22.0                        &
28.0                        \\
LI-LMC\,1522                &
0.868                       &
0.45                        &
1.05                        &
1.2                         &
                            &
                            &
                            &
1.0\rlap{0}                 &
0.9\rlap{2}                 &
$<$3.0                      &
                            \\
\hline
\end{tabular}
\end{table*}

In order to estimate bolometric luminosities and mass-loss rates, infrared
photometry was compiled to sample the spectral energy distribution (SED).
Near-IR photometry (Table 2) was retrieved from the DENIS catalogue (IJK$_{\rm
s}$ bands at 0.79, 1.23 and 2.15 $\mu$m; Cioni et al.\ 2000; cf.\ Fouqu\'{e}
et al.\ 2000) and the 2MASS catalogue (JHK$_{\rm s}$ bands at 1.24, 1.66 and
2.16 $\mu$m; cf.\ Cohen, Wheaton \& Megeath 2003). Mid-IR photometry (Table 3)
was retrieved from the MSX catalogue or from the original MSX images, from ISO
(Trams et al.\ 1999), and from IRAS scans following the procedure described in
Trams et al.\ (1999). Upper limits were determined where these would provide
useful constraints on the shape of the SED, noting that the MSX upper limits
correspond to 1-$\sigma$ levels only.

\subsection{Additional samples}

We complement our sample with objects from van Loon et al.\ (1998) and Trams
et al.\ (1999) for which spectral subclassification and mid-IR photometry is
available. No more carbon stars were included in this way, but several M-type
and two MS-type stars were added. The MS star SHV\,0524565$-$694559 was
detected in the ISO mini-survey of the LMC (Loup et al., in preparation), with
$F_{4.5}=16.2\pm2.3$ mJy and $F_{12}=5.3\pm0.7$ mJy. Another possible S star
listed in Trams et al.\ is SHV\,0522118$-$702517, but its classification is
uncertain and no subclassification is given.

A further 34 IRAS-detected objects with spectral subclassification are
available from Loup et al.\ (1997; their Table 1); these are predominantly
early-M type supergiants that are already well represented in our sample.

\section{Analysis}

\subsection{Spectral classification}

The spectral type of oxygen-rich (M-type) stars was derived from the relative
strengths of the TiO and VO molecular absorption bands, and of the Ca {\sc ii}
triplet, whilst the spectral type of carbon-rich (C-type) stars was derived
from the relative strengths of the C$_2$ and CN molecular absorption bands.
S-type stars with a carbon abundance approaching that of oxygen have prominent
ZrO bands, with MS-type stars representing a milder form of this chemical
peculiarity. For more details about these bands the reader is referred to
Turnshek et al.\ (1985).

\subsubsection{M-type stars}

%
%
\begin{figure}[]
\centerline{\psfig{figure=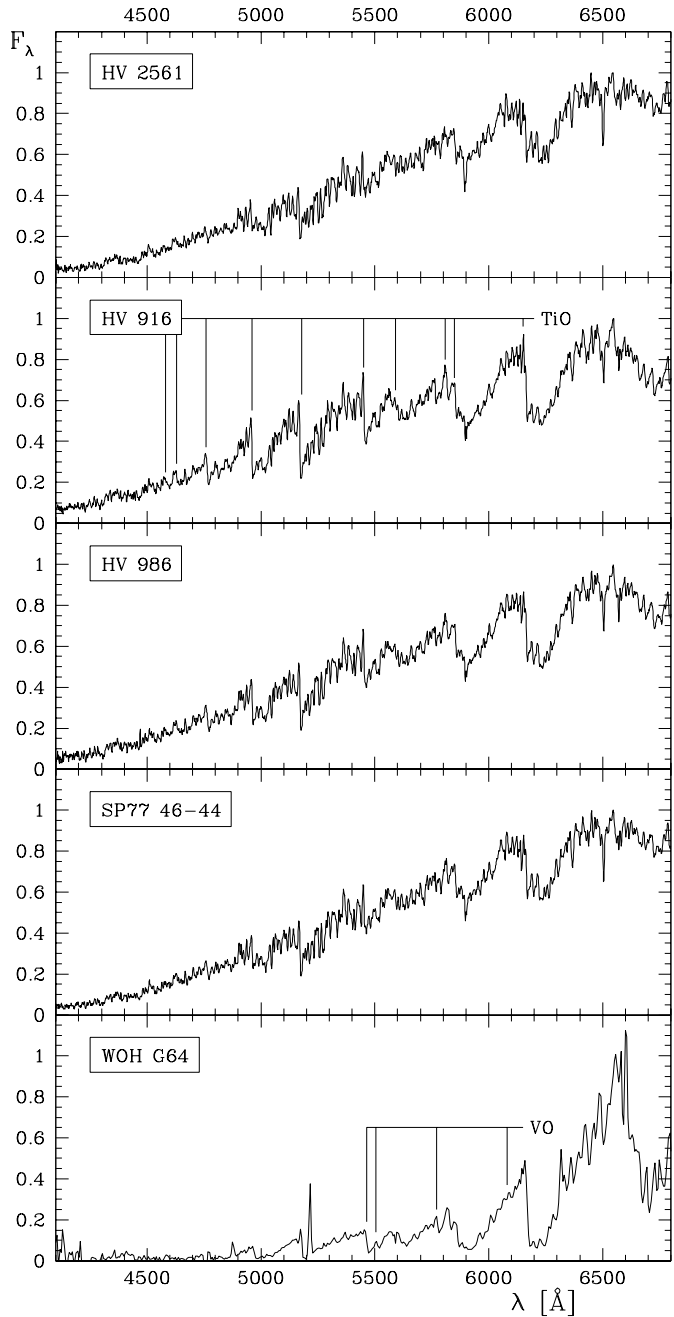,width=88mm}}
\caption[]{DFOSC spectra of M-type supergiants.}
\end{figure}

%
%
\begin{figure}[]
\centerline{\psfig{figure=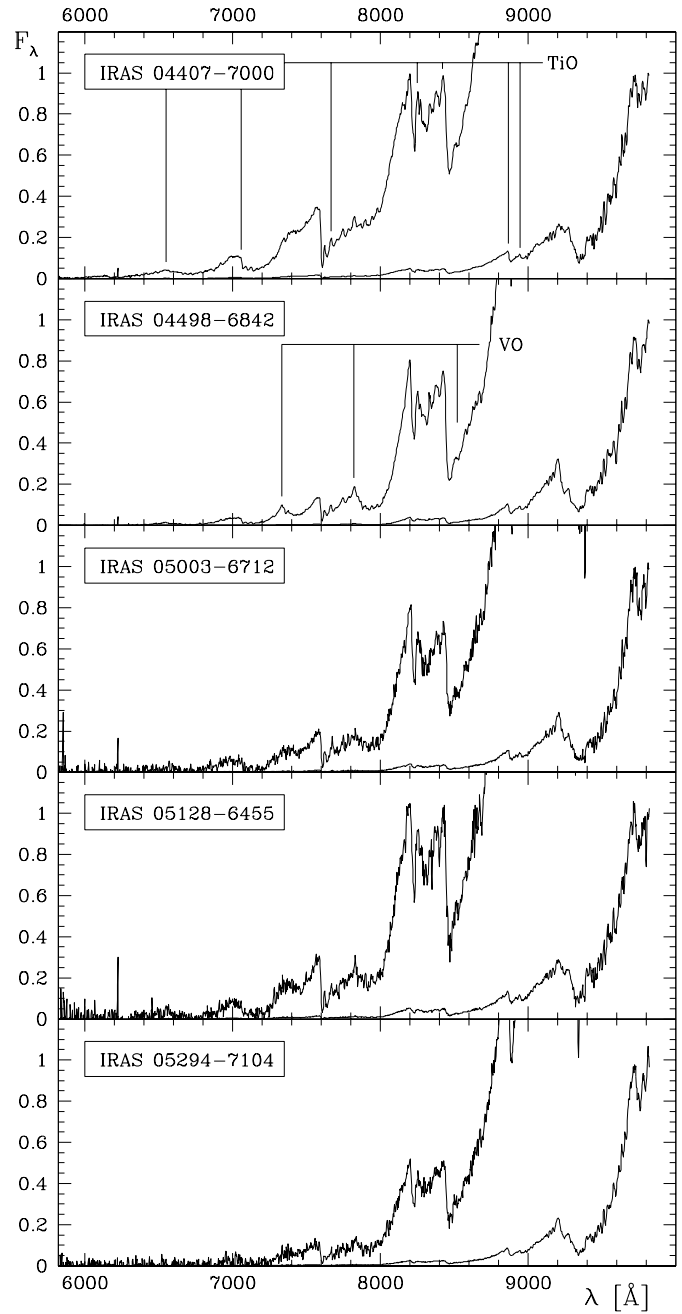,width=88mm}}
\caption[]{EMMI spectra of M-type AGB stars, displayed twice on different
scales to exploit the full dynamic range of the data.}
\end{figure}

%
%
\begin{figure}[]
\centerline{\psfig{figure=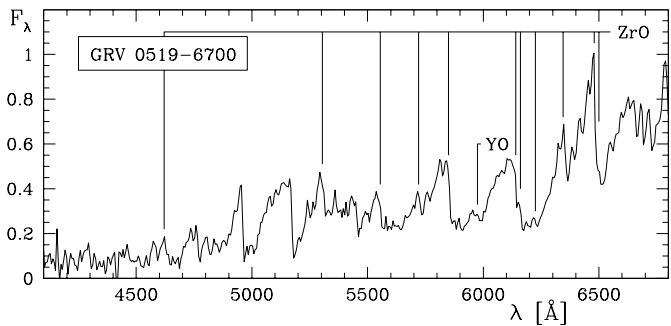,width=88mm}}
\caption[]{DFOSC spectrum of the S-type star GRV\,0519$-$6700.}
\end{figure}

%
%
\begin{figure*}[]
\centerline{\psfig{figure=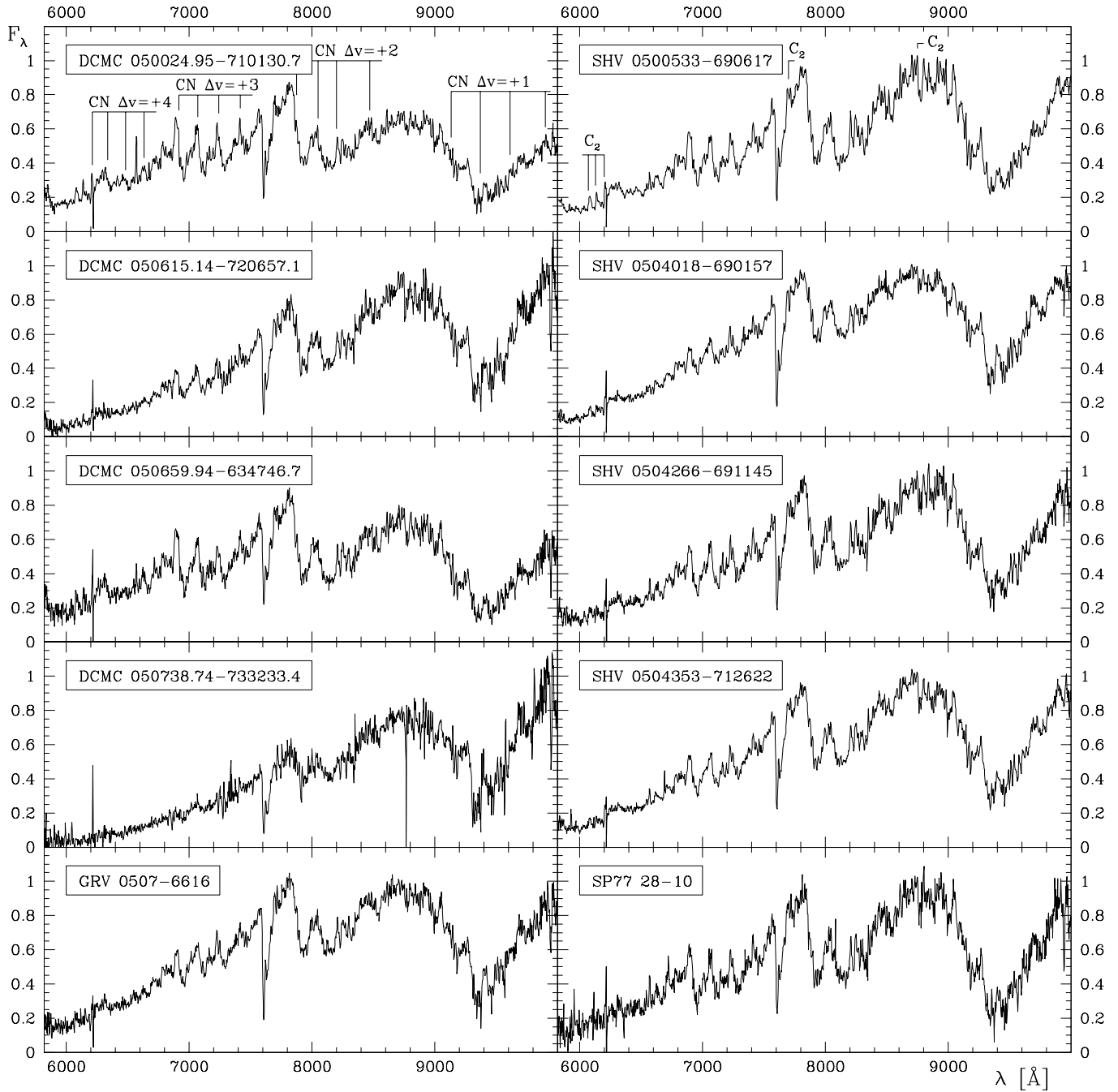,width=180mm}}
\caption[]{EMMI spectra of carbon stars.}
\end{figure*}

The M-type stars observed with DFOSC (Fig.\ 1) turn out to be red supergiants,
whilst the M-type stars observed with EMMI (Fig.\ 2) are on the AGB (see Table
4, Fig.\ 11). The 4000 to 7000 \AA\ region is not optimal for classification
of late-M types. The 7000 to 9000 \AA\ region is much better, chiefly due to
the diagnostic value of the VO bands. There can be some ambiguity, notably in
WOH\,G64 (Fig.\ 1) where various band strengths (in particular at 6200 \AA)
suggest a very late type (M7--8) whilst some relative band strengths (e.g.,
6200 compared to 5900 \AA) suggest a much earlier type ($\sim$M5). Band
strength ratios might better reflect the excitation temperature, whilst
individual bands may appear stronger as a result of enhanced optical depth in
the puffed up atmospheres of mass-losing giants. Nonetheless, normally we
obtain a spectral classification that is accurate enough for our purposes.

\subsubsection{An S-type star}

In S-type stars, the 5469 \AA\ VO band becomes weaker (e.g., relative to the
ZrO bands) with increasing C/O abundance ratio. It is quite strong in
GRV\,0519$-$6700 (Fig.\ 3), which must therefore have a C/O ratio not too
close to unity. The YO bands at 5972 and 6132 \AA, and the pair of roughly
symmetrical CaCl bands around 6200 \AA\ only appear in the cooler S-type stars
(Keenan \& Boeshaar 1980), and they are clearly present in the spectrum of
GRV\,0519$-$6700. There is no sign of $^{13}$CN bands suppressing the peak in
the spectrum between 6450 and 6456 \AA, and one can assume that this star is
not particularly $^{13}$C rich.

\subsubsection{Carbon stars}

The spectra of carbon stars (Figs.\ 4 \& 5) are generally quite similar, with
a few exceptions. The C/O abundance ratio determines the relative strength of
the C$_2$ and CN bands. The C$_2$ bands are best measured shortward of 5600
\AA, but unfortunately that spectral region was not included (and the stars
are much fainter in the V band). The best diagnostic in our spectral coverage
is the collection of C$_2$ bands shortward of 6200 \AA, and around 7700 and
8800 \AA. The temperature is best measured from the strength of the CN bands,
roughly between 6300 and 7500 \AA.

DCMC\,050738.74$-$733233.4 (Fig.\ 4) seems to be a warm carbon star and may
either be extrinsic (i.e., enriched by material from a previous AGB
companion), experiencing the effects of a thermal pulse, or on a post-AGB
track. As a consequence of the relatively high temperature all bands become
weak and it is therefore impossible to estimate the C/O ratio from our
spectrum.

IRAS\,05289$-$6617 (Fig.\ 5) exhibits extremely strong absorption in some of
the CN bands, with no sign of C$_2$. This dust-enshrouded object might be
nitrogen-enriched, possibly as a result of (mild) HBB. The mid-IR ISOCAM-CVF
spectrum (Trams et al.\ 1999) shows an emission feature around 11.5 $\mu$m
that could either be due to SiC grains or Polycyclic Aromatic Hydrocarbons
(PAHs). This star is located at the fringes of the populous elliptical
intermediate-age cluster NGC\,1978 and its colours suggest the lack of warm
dust (van Loon et al., in preparation). The latter might be explained if the
star has just recovered from a thermal pulse.

\subsubsection{Emission-line objects}

Two dusty objects turned out to exhibit a rich spectrum of strong emission
lines (Fig.\ 6) including Balmer H$\alpha$, the Ca {\sc ii} triplet and the
Paschen hydrogen series longward of 8400 \AA. There is evidence for a
two-component spectrum, with a warm (or hot) continuum peaking around 6000
\AA\ and a cold continuum dominating longward of 9000 \AA. The cool component
seems to be relatively more important in IRAS\,04530$-$6916 than in
LI-LMC\,1522.

IRAS\,04530$-$6916 has strong [S {\sc ii}] emission in the 6717+6731 \AA\
doublet. This very luminous object was interpreted as a candidate massive
Young Stellar Object on the basis of its IR colours (van Loon et al.\ 2001b).

Besides the broad H$\alpha$ line, LI-LMC\,1522 shows strong He {\sc i}
emission lines at 6678 \& 7065 \AA\ as well as emission in [Ca {\sc ii}] at
7291+7324 \AA. Strong He lines are also seen in objects such as
$\eta$\,Carinae (Wallerstein et al.\ 2001), and LI-LMC\,1522 may be an evolved
supergiant. Indeed, its spectrum resembles that of the B[e] supergiant
Hen\,S134 and like that luminous B0 star also shows the 6170 \AA\ TiO band in
emission and numerous Fe {\sc ii} emission lines in the 6000--6500 \AA\ region
(Lamers et al.\ 1998).

%
%
\begin{figure}[]
\centerline{\psfig{figure=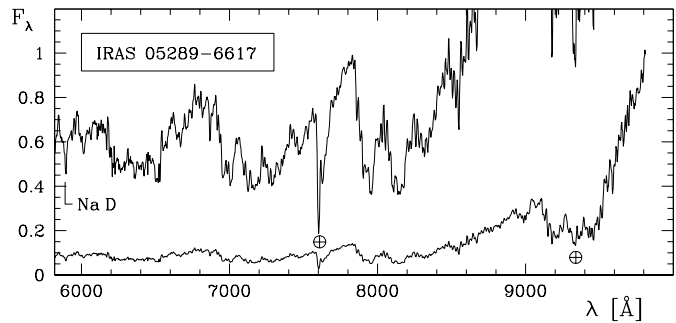,width=88mm}}
\caption[]{EMMI spectrum of the carbon star IRAS\,05289$-$6617.}
\end{figure}

%
%
\begin{figure}[]
\centerline{\psfig{figure=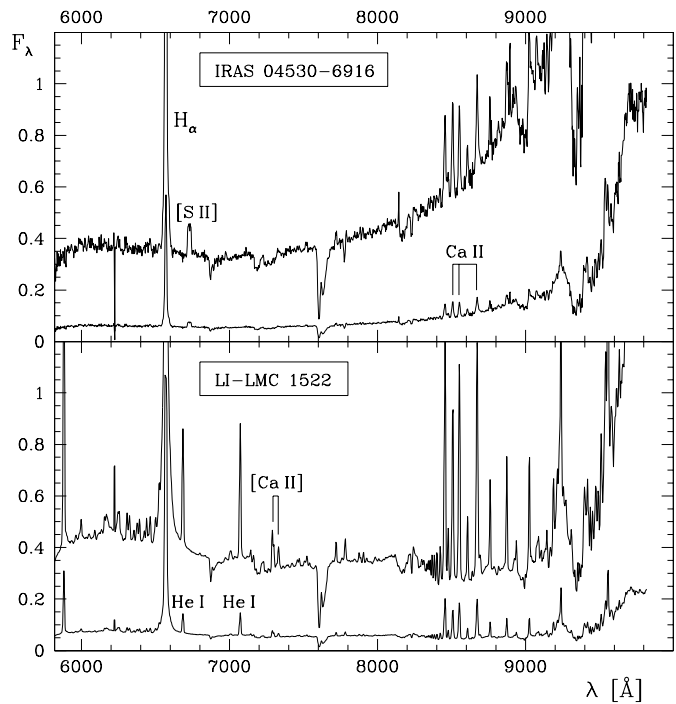,width=88mm}}
\caption[]{EMMI spectra of two dusty emission-line objects.}
\end{figure}

%
%
\begin{figure}[]
\centerline{\psfig{figure=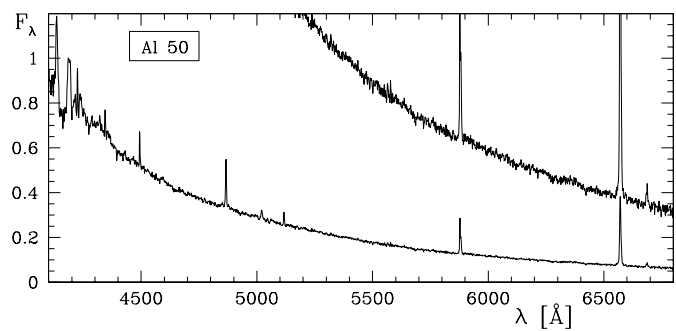,width=88mm}}
\caption[]{DFOSC spectrum of the emission-line object Al\,50.}
\end{figure}

When taking the spectrum of the dust-enshrouded AGB star IRAS\,05003$-$6712,
the B2 emission-line star Al\,50 (Andrews \& Lindsay 1964) happened to fall on
the slit of the spectrograph too. Its spectrum is characterised by bright
emission lines from the hydrogen Balmer series and helium, superimposed on a
hot continuum (Fig.\ 7).

\subsection{Modelling the spectral energy distributions}

%
%
\begin{table}
\caption[]{Results from the modelling of the spectral energy distributions:
the stellar effective temperature (from the optical spectrum), $T_{\rm eff}$,
the temperature at the inner radius of the dust envelope, $T_{\rm dust}$, the
wind speed, $v$, the bolometric luminosity, $L$, and the total (gas+dust)
mass-loss rate, $\dot{M}$. We omitted the DCMC prefix to the first four carbon
stars.}
\begin{tabular}{lccccc}
\hline\hline
Object                     &
$T_{\rm eff}$              &
\llap{$T$}$_{\rm dust}$    &
$v$                        &
\llap{$\log$} \rlap{$L$}   &
$\log \dot{M}$             \\
                           &
K                          &
K                          &
km/\rlap{s}                &
L$_\odot$                  &
\llap{M}$_\odot$/\rlap{yr} \\
\hline
\multicolumn{6}{l}{\it Stars of type M} \\
HV\,12501             &
3773                  &
800                   &
\llap{1}7.5           &
5.16                  &
$-$4.79               \\
HV\,2360              &
3736                  &
700                   &
\llap{1}5.8           &
5.11                  &
$-$4.51               \\
HV\,2446              &
3434                  &
600                   &
7.8                   &
4.35                  &
$-$5.14               \\
HV\,2561              &
3831                  &
450                   &
\llap{1}1.9           &
5.25                  &
$-$4.47               \\
HV\,5870              &
3469                  &
700                   &
\llap{1}2.7           &
4.98                  &
$-$4.68               \\
HV\,888               &
3574                  &
800                   &
\llap{1}9.7           &
5.43                  &
$-$4.43               \\
HV\,916               &
3684                  &
600                   &
\llap{1}4.5           &
5.17                  &
$-$4.50               \\
HV\,986               &
3736                  &
450                   &
9.7                   &
5.12                  &
$-$4.89               \\
HV\,996               &
3574                  &
600                   &
\llap{1}2.3           &
5.07                  &
$-$4.28               \\
IRAS\,04407$-$7000    &
3008                  &
900                   &
\llap{1}0.9           &
4.74                  &
$-$4.15               \\
IRAS\,04498$-$6842    &
2500                  &
\llap{1}100           &
\llap{1}4.4           &
4.89                  &
$-$4.30               \\
IRAS\,04509$-$6922    &
2500                  &
800                   &
\llap{1}1.7           &
4.83                  &
$-$4.45               \\
IRAS\,04516$-$6902    &
2667                  &
900                   &
\llap{1}2.6           &
4.89                  &
$-$4.23               \\
IRAS\,05003$-$6712    &
2667                  &
\llap{1}200           &
\llap{1}0.5           &
4.39                  &
$-$4.34               \\
IRAS\,05128$-$6455    &
2667                  &
\llap{1}000           &
7.2                   &
4.08                  &
$-$4.36               \\
IRAS\,05294$-$7104    &
2890                  &
900                   &
7.9                   &
4.42                  &
$-$4.14               \\
SHV\,0522023$-$701242 &
3666                  &
\llap{1}200           &
6.9                   &
3.56                  &
$-$6.80               \\
SHV\,0530323$-$702216 &
3309                  &
500                   &
4.5                   &
3.88                  &
$-$5.67               \\
SP77\,30-6            &
2890                  &
800                   &
8.5                   &
4.29                  &
$-$4.76               \\
SP77\,46-44           &
3792                  &
400                   &
\llap{1}0.9           &
5.47                  &
$-$4.56               \\
WOH\,G64              &
3008                  &
800                   &
\llap{1}5.8           &
5.69                  &
$-$3.12               \\
WOH\,SG374            &
3309                  &
600                   &
8.4                   &
4.68                  &
$-$4.43               \\
\hline
\multicolumn{6}{l}{\it Stars of type MS and S} \\
GRV\,0519$-$6700      &
3300                  &
\llap{1}200           &
8.8                   &
3.70                  &
$-$6.20               \\
HV\,12070             &
3600                  &
750                   &
\llap{1}0.8           &
4.41                  &
$-$5.20               \\
SHV\,0524565$-$694559 &
3400                  &
\llap{1}200           &
7.7                   &
3.90                  &
$-$6.53               \\
\hline
\multicolumn{6}{l}{\it Stars of type C} \\
J050024.95$-$710130.\rlap{7} &
2380                  &
800                   &
8.9                   &
4.09                  &
$-$5.05               \\
J050615.14$-$720657.\rlap{1} &
2920                  &
\llap{1}000           &
9.8                   &
4.03                  &
$-$4.96               \\
J050659.94$-$634746.\rlap{7} &
2380                  &
\llap{1}300           &
\llap{1}9.5           &
3.92                  &
$-$5.64               \\
J050738.74$-$733233.\rlap{4} &
3820                  &
\llap{1}200           &
\llap{1}4.9           &
4.05                  &
$-$5.07               \\
GRV\,0507$-$6616      &
2560                  &
\llap{1}300           &
\llap{1}7.6           &
3.77                  &
$-$5.59               \\
IRAS\,05289$-$6617    &
2200                  &
220                   &
1.5                   &
3.69                  &
$-$4.35               \\
SHV\,0500533$-$690617 &
2560                  &
900                   &
8.6                   &
3.63                  &
$-$5.39               \\
SHV\,0504018$-$690157 &
3100                  &
\llap{1}200           &
\llap{1}3.3           &
3.79                  &
$-$5.30               \\
SHV\,0504266$-$691145 &
2560                  &
\llap{1}000           &
\llap{1}0.3           &
3.62                  &
$-$5.48               \\
SHV\,0504353$-$712622 &
2740                  &
\llap{1}200           &
\llap{1}5.7           &
3.91                  &
$-$5.62               \\
SP77\,28-10           &
2380                  &
\llap{1}100           &
\llap{1}3.9           &
3.80                  &
$-$5.51               \\
\hline
\end{tabular}
\end{table}

The spectral energy distribution (SED) of each object was reproduced with the
dust radiative transfer model {\sc dusty} (Ivezi\'{c}, Nenkova \& Elitzur
1999), which was made to calculate the radial density distribution consistent
with radiation-driven wind theory (Ivezi\'{c} \& Elitzur 1995). This takes
into account the dependence of the wind speed on the stellar parameters
luminosity and metallicity (via the dust-to-gas mass ratio), and the drift
speed between the dust and the gas. This formalism is confirmed to hold for
OH/IR stars in the LMC for which the wind speeds have been measured from the
OH maser emission profiles (Marshall et al.\ 2004), a few of which are in
common with the sample we discuss here. Some of the model parameters that we
will analyse in detail are summarised in Table 4. We discuss some of the input
parameters and the reliability of the model output, before analysing the
results.

\subsubsection{Stellar effective temperatures}

The stellar effective temperatures corresponding to the M-type spectral
subclasses are taken from Fluks et al.\ (1994). These are strictly speaking
valid for luminosity class III giants of solar abundances, and differences may
be expected for metal-poor stars and for luminous supergiants. The
stratification of the extended molecular atmosphere together with variations
over the pulsation cycle limit the accuracy with which these cool giants may
be assigned a value for the effective temperature. Where more determinations
of the spectral type were available and/or classifications to half a subclass
were given, the spectral subclasses were averaged and the stellar effective
temperature was obtained by linear interpolation of the given values. The thus
adopted values are probably only accurate to $\sim$300 K (cf.\ Houdashelt et
al.\ 2000). We used the synthetic spectra from Fluks et al.\ closest to the
actual object's spectral type as models for the external radiation field when
running the {\sc dusty} code.

Despite the prominent and unique molecular absorption features characterising
carbon star spectra, it is remarkably difficult to classify these spectra and
even more difficult to assign stellar effective temperatures. We constructed
a linear grid of stellar effective temperature values between 2200 K for
spectral type C10 and 4000 K for C0. We stress that this is a purely {\it ad
hoc} temperature scale, and we will limit the quantitative interpretation of
the carbon stars in our sample, but it nevertheless allows to rank the carbon
stars according to their temperatures and study the qualitative effect that
differences in the surface temperature may have on the circumstellar envelope
and mass-loss rate. Bergeat, Knapik \& Rutily (2001) construct a {\it
photometric} temperature classification scheme, and it would be useful to
tally their findings with a {\it spectral} classification scheme. In running
{\sc dusty} we assumed a blackbody of the effective temperature as the model
for the external radiation field.

For MS and S-type stars things become even more uncertain. We assign stellar
effective temperatures in between the values we would have given were they of
either M-type or C-type. For the purpose of running {\sc dusty}, we assumed
that the spectral energy distribution of the external radiation field were that
of an equivalent M-type synthetic spectrum from Fluks et al.\ (1994).

\subsubsection{Grain properties}

Because of the limitations of the data to constrain the dust grain properties
(size, shape, composition...), we have restricted ourselves in the choice of
dust type, and assumed a grain size of 0.1 $\mu$m unless the fit of the {\sc
dusty} model to the observed SED improved notably by making a different
choice. In order to derive the total mass budget one also has to adopt a mass
density for the grains. Although this varies with the type of material the
grains are composed of, we adopt a default value of $\rho_{\rm grain}=3$ g
cm$^{-3}$.

For oxygen-rich M-type stars and MS and S-type stars we adopted the
astronomical silicate from Draine \& Lee (1984), but for the carbon stars we
took different mixtures of graphite (Draine \& Lee 1984), amorphous carbon
(Henning \& Mutschke 1997) and silicon carbide (SiC, from P\'{e}gouri\'{e}
1988). By default we used amorphous carbon only, but
DCMC\,J050738.74$-$733233.4, SHV\,0500533$-$690617 and SHV\,0504353$-$712622
required the inclusion of 10 per cent of SiC, and the ISO spectrum of
IRAS\,05289$-$6617 suggests the need of 20 percent of SiC (van Loon et al.\
1999). The SEDs of DCMC\,J050024.95$-$710130.7 and DCMC\,J050615.14$-$720657.1
are better reproduced by taking only 50 per cent of amorphous carbon, with the
remainder made up with graphite or with 40 per cent graphite and 10 per cent
SiC, respectively.

The near-IR extinction and mid-IR emission of several amongst the most
dust-enshrouded M-type stars of our sample could be better brought in line
with each other by reducing the grain size to 0.06 $\mu$m (IRAS\,5128$-$6455)
or 0.05 $\mu$m (IRAS\,04407$-$7000, IRAS\,04498$-$6842 and WOH\,G64), or by
taking a standard MRN size distribution (a power-law with exponent $-3.5$;
Mathis, Rumpl \& Nordsieck 1977) between a minimum grain size of 0.01 $\mu$m
and a maximum grain size of 0.1 $\mu$m (IRAS\,05003$-$6712). Amongst the
carbon stars we deviated from the 0.1 $\mu$m grain size only in the case of
SHV\,0504353$-$712622 (0.01 $\mu$m), and IRAS\,5289$-$6617 for which we
adopted an MRN distribution of larger grains between 0.1 and 1 $\mu$m. The
latter object has both a cool photosphere and abnormally cool dust, and with
the high fraction of SiC (see above) it is also mineralogically peculiar.

Models were first computed for a dust temperature at the inner radius of the
circumstellar envelope of $T_{\rm dust}=1000$ to 1200 K, but in many cases an
acceptable fit to the observed SED could only be achieved by lowering this
temperature significantly --- although altering the grain properties may also
have the desired effect.

\subsubsection{Luminosities}

To derive the bolometric luminosities the {\sc dusty} output SED needs to be
scaled to fit the observed SED. Adopting a distance to the LMC of 50 kpc, this
scaling factor then yields the bolometric luminosity under the assumption that
interstellar extinction is negligible over most of the infrared SED.

All objects are long-period variables. Although the amplitude of variability
drastically diminishes at IR wavelengths, it can still reach a magnitude in
bolometric luminosity for AGB stars (for supergiants the amplitude is usually
smaller in units of magnitude even if it corresponds to a larger change in
energy output). This variability is reflected in differences between the DENIS
and 2MASS near-IR magnitudes, and between the MSX, ISO and IRAS mid-IR flux
densities. We attempt to fit the {\sc dusty} model to a roughly ``mean'' SED.
Apart from the uncertainty in the distance to the LMC of around 10 per cent,
the values for the luminosities of individual stars are believed to be
accurate to within $\sim$20 per cent.

\subsubsection{Mass-loss rates}

In order to obtain total (gas+dust) mass-loss rates one has to know the
gas-to-dust mass ratio in the outflow. In the absence of knowledge thereof we
{\it assume} a value of $\psi=\rho_{\rm gas}/\rho_{\rm dust}=500$, which we
justify by scaling the typical value of $\psi=200$ for giants with solar
abundances to the lower metallicity of the LMC (cf.\ van Loon et al.\ 1999).
For most of the dust-enshrouded objects it would seem plausible that the dust
condensation process reaches its maximum efficiency, supporting these low
gas-to-dust ratios, but for less evolved and/or hotter objects this may no
longer be the case, possibly leading to underestimated values for their
mass-loss rates.

Many of the objects in our sample are detected in the IRAS data at 60 $\mu$m,
at a level much higher than can be reconciled with a constant mass-loss rate
model. Although these detections need to be confirmed at a higher sensitivity
and angular resolution, possible explanations could include the heating of
interstellar dust by the supergiants and/or ancient episodes of enhanced mass
loss. The latter is expected especially for AGB stars descending from low-mass
(1--3 M$_\odot$) Main-Sequence progenitors, as the effects of thermal pulses
on the upper AGB on the surface temperature and luminosity are much more
dramatic than for their more massive siblings (cf.\ Schr\"{o}der, Winters \&
Sedlmayr 1999).

There is ample opportunity for making different choices of grain properties,
dust temperature, optical depth etcetera without affecting the fit of the {\sc
dusty} model to the SED. This is especially true for objects with optically
thin dust shells. However, fortunately the resulting mass-loss rate is
surprisingly little influenced by this degeneracy, thanks to the scaling
inter-relationships between the quantities that govern the radiative transfer
(Elitzur \& Ivezi\'{c} 2001; cf.\ van Loon et al.\ 1997). Apart from largely
systematic uncertainties such as the gas-to-dust ratio, the values for the
mass-loss rates of individual objects are reliable to within a factor of
$\sim$two. This is confirmed by the observed spread in the $L$, $T_{\rm eff}$
dependence of the mass-loss rate (see below).

\section{Mass-loss rate as a function of bolometric luminosity and stellar
effective temperature}

\subsection{Mass-loss rates from oxygen-rich dust-enshrouded red giants in the
Large Magellanic Cloud}

%
%
\begin{figure}[]
\centerline{\psfig{figure=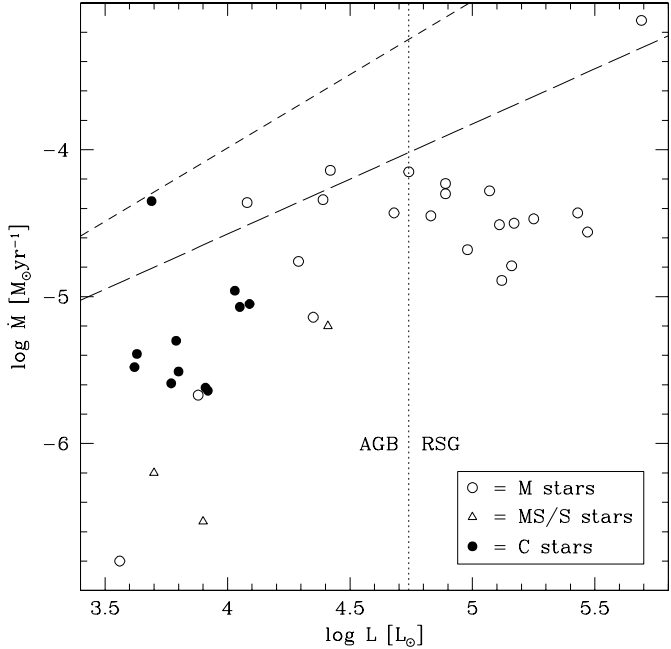,width=88mm}}
\caption[]{Mass-loss rate as a function of bolometric luminosity, for M-type
stars (circles), MS or S-type stars (triangles), and carbon stars (solid
dots). The classical AGB luminosity limit is marked by a vertical dotted line.
The long and short-dashed slanted lines mark the classical and
multiple-scattering limits to the mass-loss rate (van Loon et al.\ 1999).}
\end{figure}

The mass-loss rates reached by more luminous AGB stars are higher than those
reached by less luminous AGB stars, but there is a lot of intrinsic scatter in
the mass-loss rate versus luminosity relation due to evolutionary effects
(Fig.\ 8). Stars that have only just reached the thermal pulsing stage of AGB
evolution do not experience as high a mass-loss rate as when they reach the
AGB tip and develop a superwind. Likewise, although red supergiants feature
mass-loss rates that are comparable to those reached by luminous tip-AGB
stars, they do not seem to continue the trend of the mass-loss rate increasing
with luminosity because their true superwind stage lasts very briefly and thus
this evolutionary phase is largely missed. The one exception within our sample
is WOH\,G64, the most luminous red supergiant in the Magellanic Clouds with
the largest mass-loss rate.

Our sample is obviously biased against the most severely dust-enshrouded stars
for which no spectrum of the stellar photosphere can be taken at optical
wavelengths. This is especially true for the carbon stars, of which the
circumstellar dust envelopes become optically thick relatively easily as a
result of the higher opacity of carbon-rich dust and the more compact geometry
of the envelope (cf.\ Fig.\ 13). Hence our sample seems to have fewer examples
of the extremely high mass-loss rates as compared with the sample of obscured
objects in van Loon et al.\ (1999) and their relationships for the classical
mass-loss rate limit and mass-loss rate limit in the presence of multiple
scattering of photons off dust grains (the long and short-dashed lines in
Fig.\ 8, respectively). Nonetheless, our sample does include several OH/IR
stars which are in the superwind phase, and our optical spectroscopy of these
heavily dust-enshrouded objects makes a significant contribution to probing
the stellar parameters well into this literally obscure phase of stellar
evolution.

At least part of the evolutionary effect in the dispersion of mass-loss rates
can be related to the stellar photospheric temperature. On the AGB, stars
evolve not only to become brighter but also to become cooler, whilst red
supergiants first cool rapidly but are then expected (on the basis of models)
to increase in luminosity somewhat at an essentially constant temperature. It
is not clear at what precise evolutionary stage we capture the red supergiants
in our sample.

%
%
\begin{figure}[]
\centerline{\psfig{figure=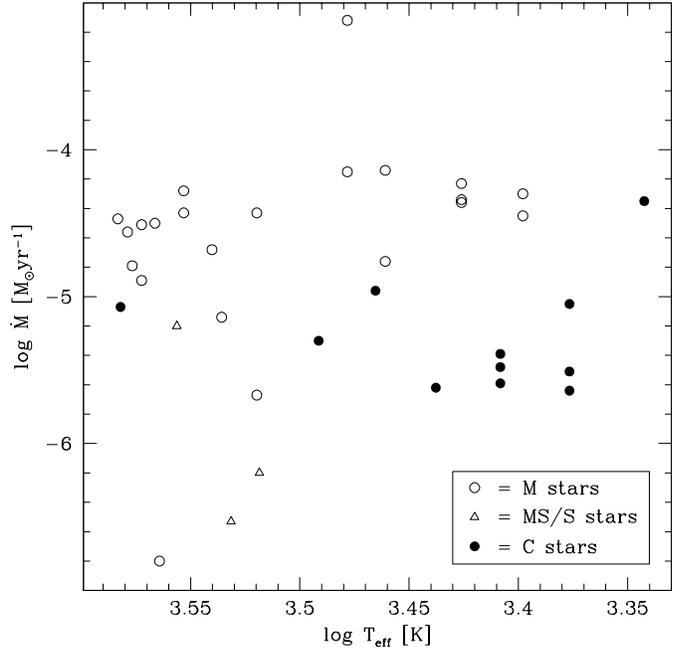,width=88mm}}
\caption[]{Mass-loss rate as a function of stellar effective temperature.
Symbols are the same as in Fig.\ 8.}
\end{figure}

When the mass-loss rate is held against the stellar effective temperature,
however, no clear trend is seen (Fig.\ 9). The M-type stars with the lowest
mass-loss rates are fairly warm and the coolest M-type stars have relatively
high mass-loss rates, but the red supergiants trouble this picture as they
have high mass-loss rates not because they are cool (they are, in fact,
amongst the warmer stars in our sample) but simply because they are luminous.
This might give the impression that the luminosity dependence of the mass-loss
rate dominates over the temperature dependence. However, it should be realised
that the range in luminosities is vastly larger (two orders of magnitude) than
the range in stellar effective temperature (less than a factor two) within our
sample.

A surface can be fit to the $\dot{M}(L,T_{\rm eff})$ plane for the M-type
stars, with useful constraints on the fit parameters. In practice, we fit a
flat surface in logarithmic space using a non-linear least-squares
Marquardt-Levenberg algorithm. We thus derive:
$$
\log \dot{M} = \alpha + \beta \log \left( \frac{L}{10,000 {\rm L}_\odot}
\right) + \gamma \log \left( \frac{T_{\rm eff}}{3500 {\rm K}} \right),
$$
with for M-type stars: $\alpha=-5.65\pm0.15$, $\beta=1.05\pm0.14$ and
$\gamma=-6.3\pm1.2$.

This solution for the mass-loss rate recipe features an approximately linear
increase of the mass-loss rate with an increase in luminosity, which is
consistent with the radiation-driven dust wind theory including
multiple-scattering (cf.\ van Loon et al.\ 1999). This is also reproduced for
a set of synthetic stars by Schr\"{o}der, Wachter \& Winters (2003), who find
a luminosity dependence with an exponent of 1.16.

Besides the luminosity dependence, the mass-loss rate is also seen to depend
sensitively on the stellar effective temperature, approximately inversely as
the sixth power. A strong temperature dependence was initially suggested by
Arndt, Fleischer \& Sedlmayr (1997) on the basis of models for carbon stars.
Although they found an even stronger dependence of the mass-loss rate on the
stellar effective temperature, with an exponent of $-8.3$, their updated model
calculations reduce the temperature exponent to $-6.8$ (Wachter et al.\ 2002);
this is fully consistent with our observed value of $-6.3\pm1.2$.

These authors also found a luminosity dependence of the mass-loss rate, with
an exponent of 1.5 (Arndt et al.\ 1997) and 2.5 (Wachter et al.\ 2002),
respectively. Although these values are significantly larger than our observed
value for the luminosity exponent of $1.05\pm0.14$ they also include a
dependence on the stellar mass, with a decreasing mass-loss rate for an
increase in stellar mass. This effect counteracts upon the luminosity
dependence and, if the mass dependence were somehow incorporated within the
luminosity dependence, it would support a lower observed value for the
luminosity exponent in the absence of an explicit parameterisation in terms of
stellar mass. Our formula may thus be consistent with the models of Arndt and
Wachter, but this must be tested with objects for which the current masses are
determined from accurate modelling of their pulsation characteristics.

%
%
\begin{figure}[]
\centerline{\psfig{figure=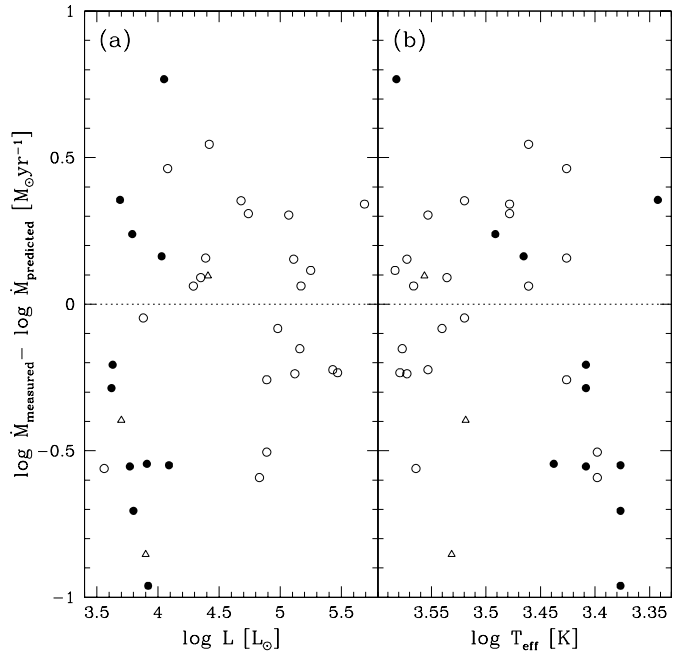,width=88mm}}
\caption[]{Discrepancies between measured mass-loss rates from modelling of
the SED, and the mass-loss rates as predicted from the $L$, $T_{\rm eff}$
dependence of the mass-loss rate for M-type stars, plotted versus bolometric
luminosity ({\bf a}) and stellar effective temperature ({\bf b}). Symbols are
the same as in Fig.\ 8.}
\end{figure}

The residuals from the $\dot{M}(L,T_{\rm eff})$ fit (Fig.\ 10) show a
dispersion of only 0.3 dex in $\log \dot{M}$ (a factor two in $\dot{M}$),
which is comparable to the accuracy of an individual estimate of the mass-loss
rate. Other stellar parameters may contribute to the spread, but it is clear
that the luminosity and temperature dominate the mass-loss rate recipe for
these stars.

We do not account for departures from spherical symmetry of the mass loss.
Asymmetric flows are often seen in Planetary Nebulae and post-RSG objects, and
thought to have their origin during the previous superwind phase. Much
amplified in post-AGB and post-RSG objects due to the interaction of a fast
wind with the slower superwind, it is not yet clear how prominent or common
these asymmetries are in the winds of AGB stars and red supergiants. There
exists no evidence for the global mass-loss {\it rate} to be affected by
asymmetry in the outflow kinematics, or by the presence of a binary companion.
Our sample might already include binaries or objects with asymmetric outflows
as we have not wittingly excluded them --- in which case the mass-loss rates
for such objects cannot be different by more than a factor two.

We now divide the sample of M-type stars into red supergiants, with
$\log{L}>4.9$, and AGB stars, with $\log{L}<4.9$ (see next section), and fit a
similar $\dot{M}(L,T_{\rm eff})$ formula to each of these subsamples. Despite
the smaller number of stars to constrain the fit (10 red supergiants and 12
AGB stars), the results are in excellent agreement with each other. In
particular, the luminosity exponent $\beta=0.82\pm0.38$ for the red
supergiants and $\beta=1.10\pm0.40$ for the AGB stars are indistinguishable.
Despite the much larger uncertainty on the temperature exponent, where we note
that the distribution of red supergiants is heavily skewed towards early
types, with $\gamma=-10.8\pm2.7$ for the red supergiants and
$\gamma=-5.2\pm3.0$ for the AGB stars these are also fully consistent with
each other. The fits lead to an estimate for the reference mass-loss rate of
$\alpha=-5.3\pm0.5$ and $\alpha=-5.6\pm0.2$ for the red supergiants and AGB
stars, respectively, again suggesting that the mass-loss rates from both red
supergiants and oxygen-rich AGB stars are governed by the same mechanism and
in the same manner.

If the mass-loss rate recipe for M-type stars is applied to the (M)S and
C-type stars, their dispersion is reduced as well, suggesting that the
mass-loss rate of these types of chemically distinct objects may depend on the
luminosity and stellar effective temperature in a similar fashion. Attempts to
fit an $\dot{M}(L,T_{\rm eff})$ relation to these classes of stars separately
lead to qualitative similar exponents (positive for the luminosity dependence
and negative for the temperature dependence) but in view of the smaller number
of stars and the uncertainty with regard to the temperature scale for these
objects we prefer not to publish the results of these fits to avoid abuse.

\subsection{Comparison with galactic red giants}

%
%
\begin{table}
\caption[]{Galactic samples of red giants (Jura \& Kleinmann 1989, 1990,
1992), with the spectral type, the stellar effective temperature, $T_{\rm
eff}$, the bolometric luminosity, $L$, and the mass-loss rates as determined
by Jura \& Kleinmann, $\dot{M}_1$, and as predicted by our recipe,
$\dot{M}_2$. We omitted the IRAS prefix to some of the dust-enshrouded AGB
stars.}
\begin{tabular}{llcccc}
\hline\hline
Object           &
Spectra\rlap{l}  &
$T_{\rm eff}$    &
$\log L$         &
$\log \dot{M}_1$ &
$\log \dot{M}_2$ \\
                 &
Type             &
K                &
L$_\odot$        &
M$_\odot$/yr     &
M$_\odot$/yr     \\
\hline
\multicolumn{6}{l}{\it Intermediate-period AGB stars (Jura \& Kleinmann
1992)} \\
S\,CrB              & M7             & 3126 & 3.70 & $-$6.96 & $-$5.66 \\
R\,Leo              & M8             & 2890 & 3.70 & $-$7.07 & $-$5.44 \\
R\,LMi              & M7             & 3126 & 3.70 & $-$6.82 & $-$5.66 \\
RX\,Boo             & M7.5           & 3008 & 3.70 & $-$6.80 & $-$5.55 \\
W\,Hya              & M7             & 3126 & 3.70 & $-$7.00 & $-$5.66 \\
\hline
\multicolumn{6}{l}{\it Dust-enshrouded AGB stars (Jura \& Kleinmann 1989)} \\
KU\,And             & M9             & 2667 & 4.00 & $-$4.72 & $-$4.91 \\
WX\,Psc             & M9             & 2667 & 4.00 & $-$4.64 & $-$4.91 \\
02316$+$6455        & M9             & 2667 & 4.00 & $-$5.07 & $-$4.91 \\
02351$-$2711        & M9             & 2667 & 4.00 & $-$5.44 & $-$4.91 \\
NML\,Tau            & M6             & 3309 & 4.00 & $-$5.22 & $-$5.50 \\
04307$+$6210        & M6             & 3309 & 4.00 & $-$5.44 & $-$5.50 \\
TX\,Cam             & M8.5           & 2778 & 4.00 & $-$5.60 & $-$5.02 \\
05411$+$6957        & M9             & 2667 & 4.00 & $-$5.23 & $-$4.91 \\
V\,Cam              & M7             & 3126 & 4.00 & $-$5.66 & $-$5.34 \\
06300$+$6058        & M9             & 2667 & 4.00 & $-$5.28 & $-$4.91 \\
GX\,Mon             & M9             & 2667 & 4.00 & $-$5.03 & $-$4.91 \\
IW\,Hya             & M9             & 2667 & 4.00 & $-$5.02 & $-$4.91 \\
V2108\,Oph          & M9             & 2667 & 4.00 & $-$5.42 & $-$4.91 \\
V774\,Sgr           & M5             & 3434 & 4.00 & $-$5.55 & $-$5.60 \\
18009$-$2019        & M8             & 2890 & 4.00 & $-$5.37 & $-$5.12 \\
18135$-$1641        & M5             & 3434 & 4.00 & $-$5.30 & $-$5.60 \\
18204$-$1344        & M8             & 2890 & 4.00 & $-$5.38 & $-$5.12 \\
V1111\,Oph          & M9             & 2667 & 4.00 & $-$5.28 & $-$4.91 \\
18413$+$1354        & M7             & 3126 & 4.00 & $-$5.26 & $-$5.34 \\
V3953\,Sgr          & M9             & 2667 & 4.00 & $-$5.17 & $-$4.91 \\
V3880\,Sgr          & M8             & 2890 & 4.00 & $-$5.00 & $-$5.12 \\
V342\,Sgr           & M9             & 2667 & 4.00 & $-$5.36 & $-$4.91 \\
20440$-$0105        & M9             & 2667 & 4.00 & $-$5.57 & $-$4.91 \\
UU\,Peg             & M7             & 3126 & 4.00 & $-$5.80 & $-$5.34 \\
RT\,Cep             & M6             & 3309 & 4.00 & $-$5.60 & $-$5.50 \\
23496$+$6131        & M9             & 2667 & 4.00 & $-$5.27 & $-$4.91 \\
\hline
\multicolumn{6}{l}{\it Red supergiants (Jura \& Kleinmann 1990)} \\
KW\,Sgr             & M0Ia           & 3895 & 5.00 & $-$5.40 & $-$4.89 \\
VX\,Sgr             & M4Ia           & 3574 & 5.18 & $-$4.40 & $-$4.47 \\
UY\,Sct             & M4Iab          & 3574 & 5.00 & $-$5.15 & $-$4.66 \\
BD$+$24$^\circ$3902 & M1Ia           & 3810 & 5.00 & $-$6.00 & $-$4.83 \\
BI\,Cyg             & M4Ia           & 3574 & 5.30 & $-$5.22 & $-$4.34 \\
KY\,Cyg             & M3.5Ia         & 3620 & 5.60 & $-$5.40 & $-$4.06 \\
NML\,Cyg            & M6III          & 3309 & 5.70 & $-$4.00 & $-$3.71 \\
$\mu$\,Cep          & M2Ia           & 3736 & 5.60 & $-$6.05 & $-$4.15 \\
PZ\,Cas             & M2Ia           & 3736 & 5.30 & $-$5.00 & $-$4.46 \\
TZ\,Cas             & M2Iab          & 3736 & 5.00 & $-$5.52 & $-$4.78 \\
S\,Per              & M3Ia           & 3666 & 5.30 & $-$5.15 & $-$4.41 \\
SU\,Per             & M3Iab          & 3666 & 5.00 & $-$5.22 & $-$4.73 \\
$\alpha$\,Ori       & M1             & 3810 & 5.00 & $-$6.70 & $-$4.83 \\
VY\,CMa             & M5I            & 3434 & 5.60 & $-$4.00 & $-$3.92 \\
EV\,Car             & M4Ia           & 3574 & 5.30 & $-$5.22 & $-$4.34 \\
CK\,Car             & M3.5Ia\rlap{b} & 3620 & 5.00 & $-$5.70 & $-$4.69 \\
IX\,Car             & M2Iab          & 3736 & 5.00 & $-$5.05 & $-$4.78 \\
V396\,Cen           & M4Ia           & 3574 & 5.30 & $-$6.00 & $-$4.34 \\
AH\,Sco             & M4             & 3574 & 5.48 & $-$5.10 & $-$4.16 \\
\hline
\end{tabular}
\end{table}

We now investigate to what extent the mass-loss rate recipe is valid for other
samples, by applying it to galactic AGB stars and red supergiants. Jura \&
Kleinmann published studies of the mass loss from red giants in the solar
neighbourhood, which provides us with three samples: (1) AGB stars pulsating
with an intermediate-period (Jura \& Kleinmann 1992), (2) dust-enshrouded AGB
stars with a large 60\,$\mu$m excess (Jura \& Kleinmann 1989), and (3) red
supergiants (Jura \& Kleinmann 1990). From their spatial distributions Jura \&
Kleinmann estimate progenitor masses of respectively 1--1.2\,M$_\odot$,
1.5\,M$_\odot$ and $>$20\,M$_\odot$ for these three samples. We selected all
M-type stars for which spectral sub-classification is available to allow a
determination of $T_{\rm eff}$, and only those stars for which Jura \&
Kleinmann estimated a mass-loss rate: 5 intermediate-period AGB stars (with
pulsation periods of 310--372\,d), 26 dust-enshrouded AGB stars and 19 red
supergiants. Their properties are listed in Table 5. Jura \& Kleinmann (1990)
list spectral types for the supergiants, but spectral types were obtained from
Simbad for the other two samples and for the red supergiant NML\,Cyg. The
values for $T_{\rm eff}$ are assigned according to the same calibration scheme
as for our sample of magellanic objects.

Jura \& Kleinmann (1992) estimated distances for the intermediate-period AGB
stars using the K-band period-luminosity relation for Mira variables. For a
few nearby examples these are consistent with determinations from Hipparcos
data (Knapp et al.\ 2003). Nevertheless, Jura \& Kleinmann {\it assumed} the
luminosities to be 5,000\,L$_\odot$. They estimated distances for the
dust-enshrouded AGB stars from the bolometric fluxes, where they {\it assumed}
that they all have a luminosity of $10^4$\,L$_\odot$ (Jura \& Kleinmann 1989).
For the supergiants they estimated distances in most part from cluster
membership (Jura \& Kleinmann 1990). The distance ranges for the samples are
80--300\,pc, 270--1100\,pc and 200--2500\,pc, respectively.

The method of Jura \& Kleinmann to derive mass-loss rates, which they applied
to all three samples in the same manner, is based on a scaling of the measured
60\,$\mu$m flux density (Jura 1986, 1987). The result depends on distance,
$d$, luminosity, $L$, and wind speed, $v$, as $\dot{M}\propto v d^2/\sqrt{L}$.
The wind speed was measured for almost every intermediate-period and
dust-enshrouded AGB star that we selected, but for only 7 out of the 19
supergiants. For the stars without a direct measurement of the wind speed they
assumed $v=15$ and 30\,km\,s$^{-1}$ for dust-enshrouded AGB stars and
supergiants, respectively. The selected intermediate-period AGB stars have a
(measured) wind speed of only $v\simeq6$\,km\,s$^{-1}$. Some supergiants also
have a slow wind, notably $\alpha$\,Ori and $\mu$\,Cep ($v=10$\,km\,s$^{-1}$)
and Jura \& Kleinmann (1990) suggest that this may be due to a much reduced
dust-to-gas ratio ($\psi\gg200$). If that is the case then their values for
the mass-loss rates of these stars will have been underestimated.

%
%
\begin{figure}[]
\centerline{\psfig{figure=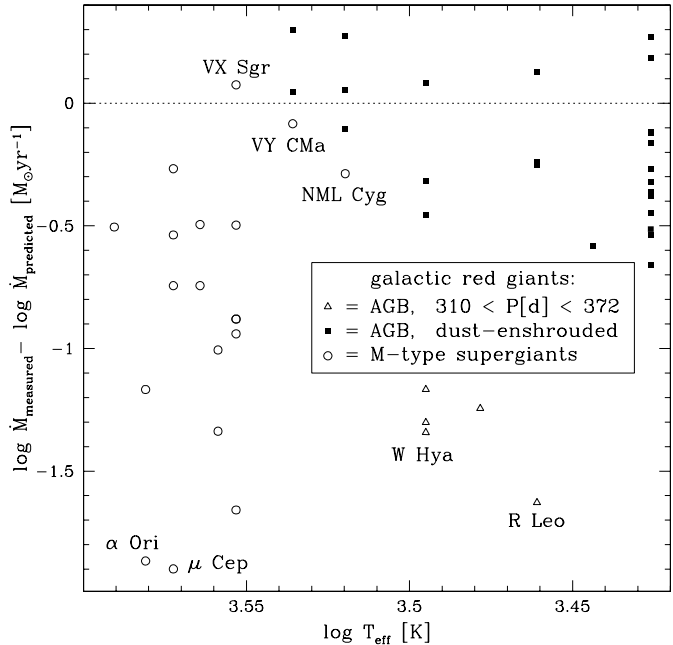,width=88mm}}
\caption[]{Discrepancies between the mass-loss rates derived from 60\,$\mu$m
flux densities, and the mass-loss rates as predicted from the
$\dot{M}(L,T_{\rm eff})$ formula for dust-enshrouded stars in the LMC, plotted
versus stellar effective temperature, for stars in the solar neighbourhood
(Jura \& Kleinmann 1989, 1990, 1992).}
\end{figure}

There is excellent agreement between the mass-loss rates as predicted using
our recipe and the measured Jura \& Kleinmann mass-loss rates for the galactic
dust-enshrouded AGB stars (Fig.\ 11, solid squares), where most of the scatter
is due to deviations from the assumed luminosity of $10^4$\,L$_\odot$ for
these stars. We note that the sample of dust-enshrouded AGB stars in Jura \&
Kleinmann (1989) includes another 5 more extreme oxygen-rich objects without
spectral sub-classification, for which they estimate mass-loss rates between
$2\times10^{-5}$ and $2\times10^{-4}$\,M$_\odot$\,yr$^{-1}$. A similar bias is
present in our magellanic sample (see next section). The good agreement
between the dust-enshrouded (oxygen-rich) AGB stars in the LMC and in the
solar neighbourhood suggests that there is no evidence for the total mass-loss
rates for these classes of stars to depend on the initial metal abundance, a
conclusion which was already reached by van Loon (2000).

Our recipe also correctly predicts the mass-loss rates for the most extreme
red supergiants, notably NML\,Cyg, VX\,Sgr and VY\,CMa that have mass-loss
rates of the order $10^{-4}$\,M$_\odot$\,yr$^{-1}$ (Fig.\ 11, circles).
However, most of the red supergiants from the Jura \& Kleinmann (1990) sample
have measured mass-loss rates that are significantly lower than the predicted
values by about a factor 3--50 (although most are within a factor 10 of the
predictions). This can be due to invalidity of the mass-loss recipe or the
assumptions made in the measured values. We already mentioned that the most
discrepant objects in our comparison, $\alpha$\,Ori and $\mu$\,Cep have
anomalously slow winds and probably a low dust-to-gas ratio. Furthermore,
these two stars are the nearest supergiants in the sample and they are
spatially resolved at 60\,$\mu$m, which may have led to a further
underestimate of the mass-loss rate. The same may be true to a lesser extent
for the other supergiants.

There is a clear discrepancy for the intermediate-period AGB stars (Fig.\ 11,
triangles), with the predictions exceeding the measured values by a factor
10--40. These are nearby objects and the 60\,$\mu$m flux may again have been
underestimated. They are cool and pulsate vigorously: 4 out of 5 are Mira-type
variables whilst W\,Hya is a semi-regular variable. But their periods are with
$P<400$\,d shorter than those of the dust-enshrouded stars, supergiants and
the magellanic sample. They are also optically bright. Our recipe is not meant
to be applied to these stars as it was derived for dust-enshrouded objects
with $P>400$\,d. The dust formation in the intermediate-period AGB stars may
not be at its maximum efficiency, resulting in lower dust-to-gas ratios
(explaining their slow winds). Perhaps the different pulsation properties or
the higher gravity near the photospheres of these stars causes the atmosphere
to be elevated less than in more extreme objects. Indeed, circumstellar dust
envelopes are much more prominent for objects with $P>400$\,d than for objects
with $P<400$\,d (e.g., Knapp et al.\ 2003).

\section{The Hertzsprung-Russell diagram of dust-enshrouded AGB stars and red
supergiants}

%
%
\begin{figure}[]
\centerline{\psfig{figure=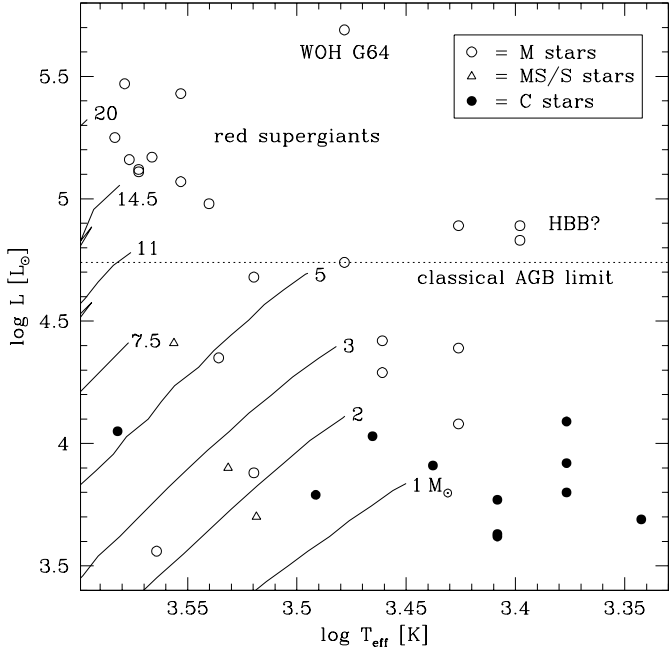,width=88mm}}
\caption[]{The Hertzsprung-Russell diagram of bolometric luminosity versus
stellar effective temperature (symbols are the same as in Fig.\ 8). The
classical AGB luminosity limit is marked by a horizontal dotted line. Three
cool stars just above this limit may be AGB stars whose luminosity is enhanced
as a result of Hot Bottom Burning (HBB). Evolutionary tracks from Bertelli et
al.\ (1994) are plotted and labelled by their Main Sequence progenitor mass.}
\end{figure}

Now that we have determined the bolometric luminosities and stellar effective
temperatures for the objects in our sample, we can place them in a physical
Hertzsprung-Russell diagram (Fig.\ 12). This allows a comparison with
theoretical evolutionary tracks to be made. To this purpose, the tracks from
Bertelli et al.\ (1994) are plotted, for a typical LMC metallicity of
[Fe/H]=$-0.4$.

Many of the stars in our sample are cooler than the endpoints on the AGB or
RSG tracks. Indeed, despite the low metal abundance in the LMC, very late
spectral types are encountered, with now at least two M10 stars known on the
AGB and for instance the red supergiant WOH\,G64 also being much cooler than
the evolutionary models produce (cf.\ Massey \& Olsen 2003). The late-type AGB
stars in our sample are not even the most extreme objects; more heavily
obscured OH/IR stars such as IRAS\,05298$-$6957 (Wood et al.\ 1992; van Loon
et al.\ 2001a) exist, for which no optical spectrum has yet been possible to
take.

On the basis of the evolutionary tracks, the classical AGB luminosity limit
and the distribution of the stars in the diagram, it seems {\it a-posteriori}
fairly obvious which stars are intermediate-mass AGB stars and which are more
massive supergiants. Rather than interpreting the three cool stars just above
the classical AGB limit (Fig.\ 12) as much cooler than expected supergiants,
it seems more logical to interpret these as slightly overluminous tip-AGB
stars as there is an accepted physical mechanism for doing so: Hot Bottom
Burning (HBB). The extra luminosity is generated from nuclear burning at the
bottom of the deep convective zone in massive ($\gsim$4 M$_\odot$) AGB stars
(Boothroyd \& Sackmann 1992).

The location of the (M)S-type and carbon stars is as expected. Note that the
Bertelli et al.\ (1994) tracks do not account for the molecular opacity efects
on the atmospheres of carbon stars which are, however, incorporated in more
recent models by Marigo (2002). We should also emphasize that especially the
estimated temperatures of the (M)S and C-type stars are uncertain in a
systematic way, and that in particular the positions of the warmest MS-type
star (HV\,12070) and the warmest carbon star (DCMC\,J050738.74$-$733233.4) in
the diagram may need to be corrected towards lower temperatures --- otherwise
the implied Main-Sequence progenitor mass of 5--6 M$_\odot$ seems inexplicably
large. These two objects have spectra that are consistent with a C/O ratio
(very) close to unity which would cause a reduced opacity in the photosphere.

Notwithstanding this, it is clear from the diagram (Fig.\ 12) that the carbon
stars in our sample are descended from low-mass Main Sequence stars (mostly in
the range of roughly 1--2 M$_\odot$) whereas the M-type AGB stars have more
massive progenitors (typically 3--5 M$_\odot$). This is as expected, as HBB
prevents the formation of massive carbon stars (Boothroyd \& Sackmann 1993;
cf.\ van Loon et al.\ 2001a; Bergeat, Knapik \& Rutily 2002; Marigo, Girardi
\& Chiosi 2003). S-type stars (including stars of MS or CS-type) are thought
to represent an intermediate phase in the formation of a carbon star from an
oxygen-rich star earlier on the AGB (Lloyd Evans 1983). It is therefore not
surprising that the (M)S-type stars in our sample are relatively warm and
similarly luminous as the carbon stars (i.e.\ fainter than most M-type stars
in our sample).

%
%
\begin{figure}[]
\centerline{\psfig{figure=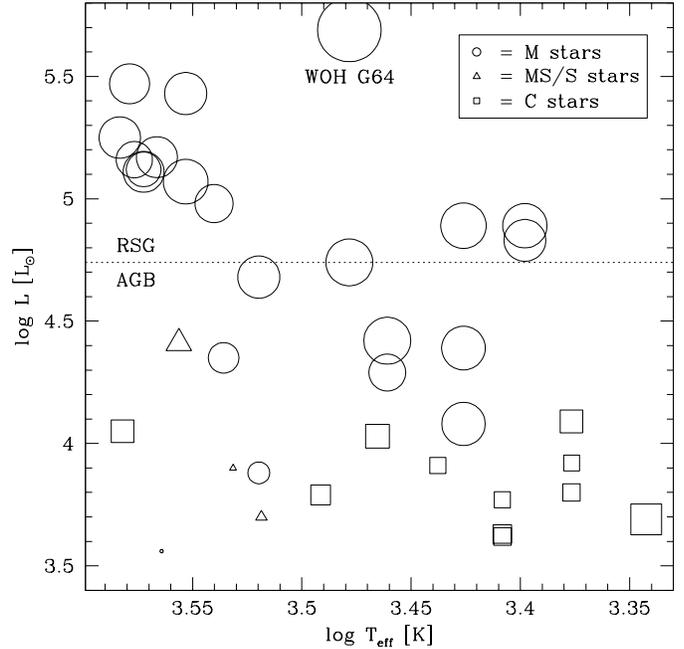,width=88mm}}
\caption[]{The Hertzsprung-Russell diagram for the M-type stars (circles), MS
or S-type stars (triangles), and carbon stars (squares), where the sizes of
the symbols are logarithmically proportional to the mass-loss rate.}
\end{figure}

The mass-loss rate can be seen to vary across the Hertzsprung-Russell diagram
due to the evolution of the stars and the dependence of the mass-loss rate on
luminosity and stellar effective temperature (Fig.\ 13). The (M)S-type stars
and the relatively warm and faint M-type AGB stars have the lowest mass-loss
rates in this diagram, whilst the coolest and most luminous carbon stars and
especially M-type AGB stars and red supergiants experience the most intense
mass loss. This illustrates the validity of the $\dot{M}(L,T_{\rm eff})$
recipe we derived earlier from these data.

%
%
\begin{figure}[]
\centerline{\psfig{figure=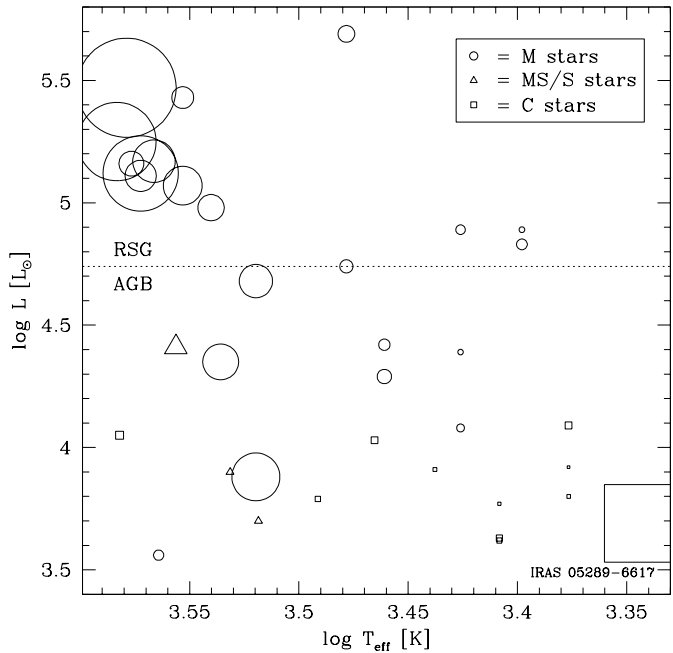,width=88mm}}
\caption[]{Same as Fig.\ 13, but now the sizes of the symbols are proportional
to $(T_{\rm eff}/T_{\rm dust})^2$, which is a measure of the distance to the
star of the dust formation region, $R_{\rm dust}/R_\star$.}
\end{figure}

There is a clear tendency for the dust shells to be warmer for cooler stellar
photospheres (Fig.\ 14). Although this may first seem odd, it can probably be
understood by the dust formation region moving in closer to the star as the
envelope becomes denser (or {\it vice versa}): the dust formation radius,
$R_{\rm dust}$, expressed in stellar radii, $R_\star$, is related to the dust
temperature at the inner edge of the circumstellar dust envelope as $R_{\rm
dust}/R_\star\propto(T_{\rm eff}/T_{\rm dust})^2$. The objects with the
relatively coldest dust and largest dust-free inner cavity are the early-M
type supergiants and the coolest and most obscured carbon star in our sample,
IRAS\,05289$-$6617 ($T_{\rm eff}\simeq2200$ K and $T_{\rm dust}\simeq220$ K).
The latter may have temporarily reduced its mass-loss rate as a result of a
thermal pulse or (less likely given the short timescale) if it has just become
a post-AGB object.

The fact that the warm and dense circumstellar dust envelopes surround the
objects with the larger mass-loss rates and cooler stellar photospheric
temperatures is not surprising, as both the mass-loss rate and the cooling of
the stellar photosphere result from the elevation of the atmospheric layers of
the star due to the strong radial pulsations which develop on the thermal
pulsing AGB and during the red supergiant stage. This may explain the
remarkable strength of the molecular bands seen in the optical spectra of
dust-enshrouded M-type AGB stars as well as in the optical and near-IR spectra
of dust-enshrouded carbon stars in the LMC (van Loon, Zijlstra \& Groenewegen
1999; Matsuura et al.\ 2002, 2005).

\section{Summary}

We presented new optical spectra for a sample of red supergiants and both
oxygen-rich and carbon-rich dust-enshrouded AGB stars in the LMC. Some of
these AGB stars have extremely cool stellar photospheres, with spectral types
as late as M10, despite their low metallicity. This is cooler than expected on
the basis of stellar evolutionary models. The same is found for the red
supergiants.

We complemented these new temperature determinations with similar LMC objects
from the literature for which spectral subclassifications are available, as
well as with infrared photometry to sample their spectral energy distributions
that were then modelled with the radiative transfer code {\sc dusty} to
measure the mass-loss rates. The mass-loss rates were found to increase with
increasing luminosity and decreasing stellar effective temperature. For the
oxygen-rich, M-type AGB stars and red supergiants these dependencies were
quantified by the mass-loss rate formula $\log \dot{M} = -5.65 + 1.05 \log ( L
/ 10,000\, {\rm L}_\odot ) -6.3 \log ( T_{\rm eff} / 3500\, {\rm K} )$, with
no evidence for a difference between the red supergiants and oxygen-rich AGB
stars. This formula prescribes the mass-loss rate to within a factor two for
oxygen-rich stars near the tip of their AGB evolution or during the red
supergiant phase when they experience strong and regular pulsation and
enshroud themselves in dusty envelopes, and is valid in particular for the
superwind phase of extreme mass loss. It thus complements the large body of
existing formulae derived for less massive and less evolved red giants, and
can be used in models for stellar and galactic evolution in stead of
extra-polations of those formulae.

Application of our formula to samples of galactic oxygen-rich red giants shows
excellent agreement between our predictions and mass-loss rates derived from
60\,$\mu$m flux densities for dust-enshrouded AGB stars and the most extreme
red supergiants. This suggests that the formula is valid for stars with masses
from $M_{\rm ZAMS}\sim$1.5\,M$_\odot$ to $M_{\rm ZAMS}>$20\,M$_\odot$, and
that the total (gas+dust) mass-loss rate does not depend sensitively on the
initial metal abundance. On the other hand, our formula predicts higher rates
than the observational estimates for less extreme red giants. This failure may
indicate that our formula should only be applied to {\it dust-enshrouded}
objects where the dust condensation efficiency is maximal, but it may also be
explained by errors in the observational estimates if the dust-to-gas ratio in
these less extreme objects is lower than what is commonly assumed.

Carbon stars and stars of type MS or S seem to show a similar behaviour of the
mass-loss rate as a function of luminosity and temperature, but a
quantification of these dependencies awaits a more careful and extensive
analysis of these classes of objects.

We presented Hertzsprung-Russell diagrams, which not only illustrate the trend
of mass-loss rate with luminosity and temperature but also a trend for dust to
form at relatively greater distances from early-M type stars than in the case
of stars with cooler stellar photospheres. These diagrams also showed evidence
for Hot Bottom Burning preventing carbon star formation for massive AGB stars
and enhancing their luminosity at the tip of the AGB.

\begin{acknowledgements}
We are indebted to ESO for the generous allocation of Director's Discretionary
Time to take the 1996 and 1998 NTT/EMMI spectra. We would like to thank the
anonymous referee for helping us clarify the presentation of our manuscript
and for encouraging us to test our mass-loss rate recipe on galactic stars.
This research has made use of the SIMBAD database, operated at CDS,
Strasbourg, France.
\end{acknowledgements}

\appendix

\section{Spectra of galactic M-type stars}

%
%
\begin{table*}
\caption[]{Spectroscopic targets in the Milky Way, listed in order of
increasing Right Ascension (J2000 coordinates) and accompanied by the
observing run code (either ``D5'' = DFOSC 1995, or ``E0'' = EMMI 2000). Also
listed are their spectral classifications and near-IR magnitudes and mid-IR
flux densities (in Jy, with the wavelengths in $\mu$m).}
\begin{tabular}{lccr|ll|rrr|rrrr}
\hline\hline
Object      &
RA          &
Dec         &
\llap{R}un  &
\multicolumn{2}{|c}{Spectral type} &
\multicolumn{3}{|c}{2MASS}  &
\multicolumn{4}{|c}{IRAS}    \\
                            &
                            &
                            &
                            &
Here                        &
Previous                    &
1.24                        &
1.66                        &
2.16                        &
12                          &
25                          &
60                          &
100                         \\
\hline
DI\,Eri       &
2 37 46.8     &
$-$45 37 10   &
D5            &
M6            &
M5/6\,III     &
4.41          &
3.37          &
3.12          &
3.1\rlap{5}   &
0.8\rlap{5}   &
0.2           &
$<$0.3        \\
DG\,Eri       &
4 20 41.3     &
$-$16 49 48   &
D5            &
M4            &
M3\,III       &
2.99          &
2.11          &
1.85          &
8.7           &
2.3\rlap{7}   &
0.4\rlap{3}   &
$<$0.4        \\
R\,Dor        &
4 36 45.5     &
$-$62 04 38   &
D5            &
M7.5          &
M8\,IIIe      &
$-$2.65       &
$-$3.73       &
$-$4.23       &
\llap{5}190.0 &
\llap{1}500.0 &
255.0         &
79.0          \\
SY\,Men       &
4 39 16.1     &
$-$74 21 42   &
E0            &
M10           &
M             &
4.89          &
3.59          &
2.81          &
51.0          &
46.0          &
7.4           &
2.1           \\
T\,Lep        &
5 04 50.8     &
$-$21 54 16   &
D5            &
M6.5e         &
M6e,var       &
1.19          &
0.20          &
$-$0.27       &
162.0         &
58.0          &
9.8           &
3.5           \\
RS\,Vel       &
9 23 45.3     &
$-$48 52 00   &
D5            &
M10e          &
M7e           &
1.29          &
0.37          &
$-$0.18       &
197.0         &
84.0          &
15.9          &
4.0\rlap{?}   \\
\hline
\end{tabular}
\end{table*}

During the DFOSC 1995 and EMMI 2000 runs, spectra were taken of several
galactic M-type stars for reference purposes. The observations were performed
and reduced in exactly the same way as the LMC targets, but with exposure
times ranging between only 5 seconds and 5 minutes. The spectral
classification and IR photometry from 2MASS and IRAS are listed in Table A1,
and the spectra are displayed in Figs.\ A1 \& A2. RS\,Vel was also covered by
MSX, yielding flux densities of $F_{8.28}=266.1$, $F_{12.13}=244.6$,
$F_{14.65}=188.3$ and $F_{21.34}=121.5$ Jy.

%
%
\begin{figure}[]
\centerline{\psfig{figure=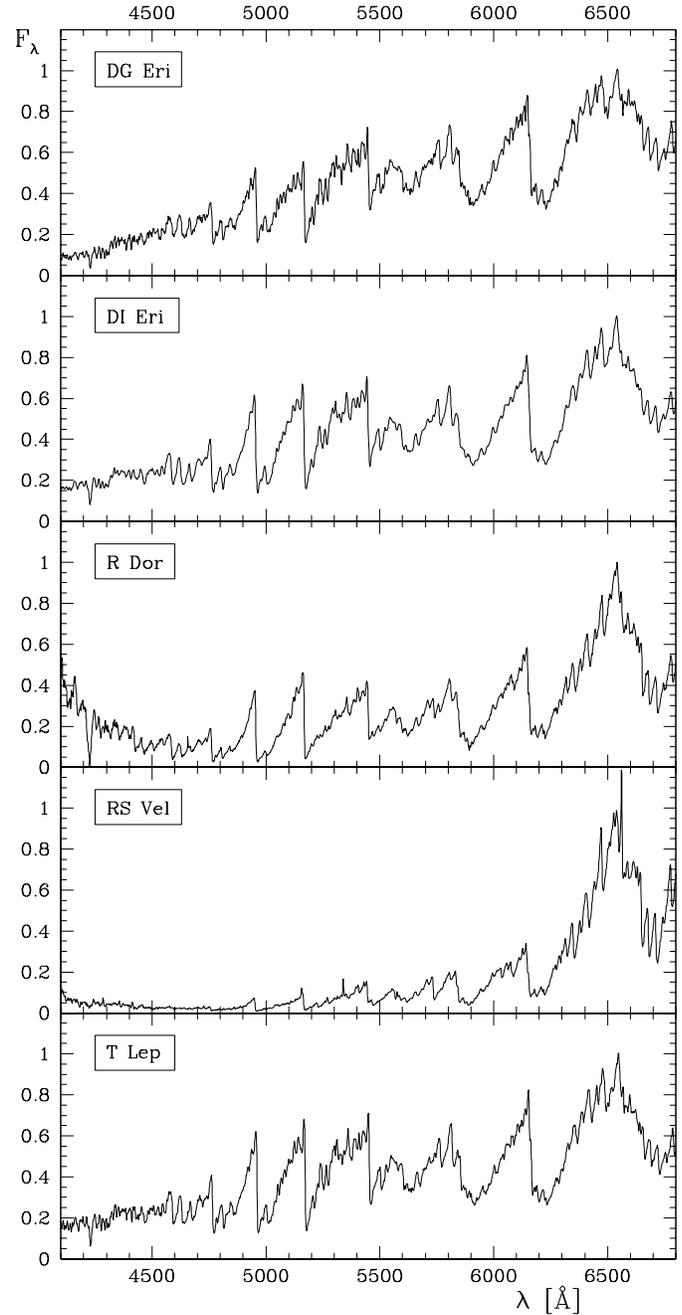,width=88mm}}
\caption[]{DFOSC spectra of galactic M-type stars.}
\end{figure}

%
%
\begin{figure}[]
\centerline{\psfig{figure=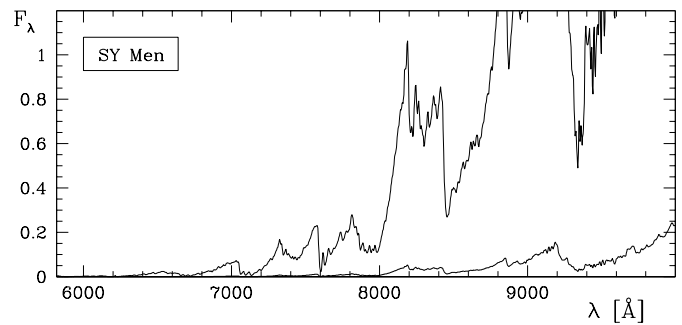,width=88mm}}
\caption[]{EMMI spectrum of the galactic M-type star SY\,Men.}
\end{figure}

\end{document}